\documentclass[useAMS,usenatbib]{mn2e}

\usepackage{ulem}
\usepackage{cancel}
\usepackage{xcolor}
\usepackage{amsmath}
\usepackage{amssymb}
\usepackage{graphicx}
\usepackage{url}
\usepackage{comment}
\usepackage{natbib}
\usepackage{fixltx2e}
\usepackage{caption}
\usepackage{subcaption}
\usepackage{soul}
\newcommand\sect{Section}

\newcommand\tabl{Table}

\newcommand\casa{\textsl{CASA}~}

\newcommand {\apgt} {\ {\raise-.5ex\hbox{$\buildrel>\over\sim$}}\ }
\newcommand {\aplt} {\ {\raise-.5ex\hbox{$\buildrel<\over\sim$}}\ } 

\newcommand\rplus{}

\title[Diffuse emission in cool-core clusters]{Mpc-scale diffuse radio
  emission in two massive cool-core clusters of galaxies}

\author[Sommer et al.]{Martin~W.~Sommer$^{(1)}$\thanks{E-mail:
    mnord@astro.uni-bonn.de (MWS); kbasu@astro.uni-bonn.de (KB)},
  Kaustuv Basu$^{(1)}$, Huib Intema$^{(2)}$, Florian Pacaud$^{(1)}$,
  \and Annalisa Bonafede$^{(3)}$, Arif Babul$^{(4,5)}$, Frank Bertoldi$^{(1)}$
  \\ \\
  (1) Argelander-Institut f\"{u}r Astronomie, Auf dem H\"ugel
  71,D-53121
  Bonn, Germany \\
  (2) National Radio Astronomy Observatory, 1003 Lopezville Road,
  Socorro, NM
  87801-0387, USA \\
  (3) Hamburger Sternwarte, Universit\"at Hamburg, Gojenbergsweg 112, 21029,
Hamburg, Germany \\
  (4) Department of Physics and Astronomy, University of Victoria, Victoria BC V7G 2H1, Canada \\
  (5) Center for Theoretical Astrophysics and Cosmology, Institute for
  Computational Science, University of Zurich, \\ \hspace{0.38cm} Winterthurerstrasse
  190, CH-8057 Z\"{u}rich, Switzerland}

\begin{document}

\date{MNRAS submitted}


\maketitle

\label{firstpage}

\begin{abstract}
  Radio halos are diffuse synchrotron sources on scales of $\sim$1 Mpc
  that are found in merging clusters of galaxies, and are believed to
  be powered by electrons re-accelerated by the merger-driven
  turbulence. We present measurements of extended radio emission on
  similarly large scales in two clusters of galaxies hosting cool
  cores: Abell 2390 and Abell 2261. The analysis is based on
  interferometric imaging with the JVLA, VLA and GMRT. We present
  detailed radio images of the targets, subtract the compact emission
  components, and measure the spectral indices for the diffuse
  components. The radio emission in A2390 extends beyond a known
  sloshing-like brightness discontinuity, and has a very steep
  \rplus{in-band} spectral slope at 1.5 GHz that is similar to some
  known ultra-steep spectrum radio halos. The diffuse signal in A2261
  is more extended than in A2390 but has lower luminosity. X-ray
  morphological indicators, derived from XMM-Newton X-ray data, place
  these clusters in the category of relaxed or regular systems,
  although some asymmetric features that can indicate past minor
  mergers are seen in the X-ray brightness images. If these two
  Mpc-scale radio sources are categorized as giant radio halos, they
  question the common assumption of radio halos occurring exclusively
  in clusters undergoing violent merging activity, in addition to
  commonly used criteria in distinguishing between radio halos and
  mini-halos.
\end{abstract}


\section{Introduction}
\label{sec:intro}

The intra-cluster medium (ICM) in massive clusters of galaxies is host
to various populations of ultra-relativistic particles (cosmic rays)
that show their presence mainly through synchrotron emission at radio
wavelengths. While some of the synchrotron emission is associated with
radio galaxies (such as radio jets and radio lobes), there is a clear
distinction from the more diffuse components observed in the direction
of some clusters. \textit{Radio relics} are one class of diffuse
sources that are generally associated with shocks originating in
merger events \rplus{\citep{1998A&A...332..395E}}, and are typically
observed at the outskirts of clusters. \textit{Radio halos} (also
\textit{Giant radio halos}, or GRHs), typically on the scale of 1 Mpc
and located in the cluster inner regions, are a class of diffuse
sources that indicate the presence of cluster-wide particle
acceleration mechanisms. \textit{Radio mini-halos} are another class
of diffuse synchrotron sources that have been classified as distinct
from GRHs \rplus{(See, e.g., \cite{2012A&ARv..20...54F} for a
  review)}.

Giant Radio halos have preferentially been found in dynamically
disturbed systems, provisionally indicating a turbulent origin of the
cosmic ray acceleration
\citep{2001ApJ...553L..15B,2010ApJ...721L..82C}. One striking property
of radio halos is a bi-modal distribution
\rplus{\citep{2007ApJ...670L...5B}}, where\rplus{by} galaxy clusters
appear divided into on- and off-states in the radio regime, although
it has been cautioned that such a bi-modality may also come about by
selection biases \citep{2012MNRAS.421L.112B}. Furthermore, the
relative prevalence of radio halos is still largely
unknown. \cite{2014MNRAS.437.2163S} found that the selection of galaxy
clusters from X-ray surveys biases against disturbed systems, thereby
underestimating the fraction of radio halo clusters.
\cite{2015A&A...580A..97C} found a large fraction of radio halos in
high-mass clusters, where most of them conform to the criteria of
merging systems, following the morphological classification discussed
in \citet{2010ApJ...721L..82C}.

Radio mini-halos have been classified as distinct from GRHs mainly due
to their compact sizes ($\sim$100$-$200 kpc) and their occurrence in
the central regions of cool-core clusters
\citep[e.g.,][]{2002A&A...386..456G,2004A&A...417....1G}. The number
of known mini-halos is small compared to known GRHs
\rplus{\citep{2014ApJ...781....9G,2015fers.confE..47B}}, and their
production mechanism is even more uncertain than that of the GRHs. One
suggested explanation invokes the mechanism of particle acceleration
through ICM turbulence as in the GRHs, but with the source of
turbulence being gas sloshing at the cluster core
\citep[e.g.,][]{2008ApJ...675L...9M,2013ApJ...762...78Z}.
Alternatively, a hadronic origin of the mini-halos, in which
radio-emitting electrons are generated from interactions between
cosmic-ray protons and the thermal ICM protons, has been proposed
\citep{2013MNRAS.428..599F,2014MNRAS.438..124Z}.

In the framework of turbulent models, the discovery of a giant radio
halo in a cool core cluster can be considered surprising, since cool
cores are generally associated with relaxed objects.  The elevated
level of turbulent kinetic energy that can drive particle acceleration
is only sustainable on timescales of the order $\lesssim 1$ Gyr
\citep{2011MNRAS.418.2467H,2014ApJ...782...21M}, within which a
disrupted cool core will not be able to re-assemble. In spite of this
theoretical expectation, a first identification of a GRH in a
cool-core cluster was recently made \citep{2014MNRAS.444L..44B}.  In
this work, we present the analysis of extended, Mpc scale diffuse
radio emissions in the two galaxy clusters Abell 2390 and Abell
2261. Neither of these systems has previously been classified as a
merging system, and \cite{2000MNRAS.315..269A} found both objects to
host cool cores (More recent work by \citealt{2013ApJ...767..116M}
also classify both objects as cool-core systems, although the
classification of A2261 can be said to be marginal). We point out that
central diffuse radio emissions has been previously identified in both
clusters, although neither was considered as hosting a giant radio
halo. In this work, we show the presence of diffuse radio emission at
a scale of 1 Mpc in these objects, based on observations using JVLA,
VLA and GMRT. \rplus{To inverstigate the dynamical states of the
  systems, we use X-ray photometry from XMM-Newton.}

Abell 2390 (henceforth A2390) is a massive cool-core cluster at
 \rplus{$z$=}0.228 \rplus{(redshift from
  \citealt{1999ApJS..125...35S})}, previously identified to host a
radio mini-halo by \cite{2003A&A...400..465B}. We re-analyze the 1.4
GHz VLA data originally presented by Bacchi et al. In addition, we
present more recent observations with the JVLA at the same frequency,
allowing for a deeper image with greater spatial dynamic range and the
determination of the spectral slope of the diffuse component.

\cite{2008A&A...484..327V} identified hints of diffuse emission in
Abell 2261 (henceforth A2261), another massive cool-core cluster at
 \rplus{$z$=}0.224 \rplus{(redshift from
  \citealt{1999ApJS..125...35S})}, from observations with the VLA. We
re-analyze the relevant archived data, and use further archived VLA
data with higher spatial resolution to model and subtract the compact
emission. Our analysis confirms the presence of diffuse Mpc-scale
radio emission in this galaxy cluster. 

We outline the interferometric \rplus{and X-ray} observations in
Section~\ref{sec:obs}, and describe our data analysis in
Section~\ref{sec:meth}. Our main results are presented in
Section~\ref{sec:res}. We discuss our findings in
Section~\ref{sec:discussion} and offer our conclusions in
Section~\ref{sec:conclusions}. For all results derived in this work we
assume a $\Lambda$CDM concordance cosmology with $h=0.7$, $\Omega_{m}
h^2 = 0.13$ and $\Omega_{\Lambda}=0.74$. Given the similar redshifts
of the two clusters, the angular to physical scale conversion is then
roughly 200 kpc/arcmin for both targets.

\section{Targets and observations}
\label{sec:obs}

The two clusters we focus on in this work, A2390 and A2261, were
identified by analyzing the radio data of a sample of 26 galaxy
clusters, selected to be complete above a mass threshold in the
redshift range $0.2 \leq z \leq 0.4$. Masses were inferred from the
integrated Comptonization of the clusters, as measured from their
Sunyaev-Zel'dovich (SZ) effect in the ESZ Planck catalog
\citep{2011A&A...536A...8P}.  In addition, independent estimates of
the masses are available from the weak-lensing analysis of the
\rplus{Canadian Cluster Comparison Project (CCCP; }
\citealt{2015MNRAS.449..685H}), with $M_{500}$ determined from a
\rplus{Navarro-Frenk-White (}NFW) fit to be $14.2^{+2.4}_{-2.3}$ and
$15.2^{+2.6}_{-2.5} \times 10^{14} h_{70}^{-1} M_{\odot}$ for A2390
and A2261, respectively. These two targets are thus some of the most
massive cool core clusters known.

It is known that X-ray selected samples are biased towards cool core
clusters \citep[e.g.,][]{2011A&A...526A..79E}, whereas an SZ
selection is less sensitive to cluster dynamical states and provides a
better mass proxy than the X-ray luminosity
\citep[e.g.][]{2005ApJ...623L..63M}. In \citet{2014MNRAS.437.2163S} we
argued that this can result in an under-representation of giant radio
halos in X-ray selected samples. However, the two present target
clusters are massive enough to be unaffected by such selection
considerations; indeed both A2390 and A2261 are also part of the GMRT
radio halo survey \citep{2015A&A...579A..92K} which is X-ray
selected. The SZ-based selection thus does not play a major role in
this work apart from sorting out some of the most massive clusters for
radio follow-up studies.

We primarily made use of our proprietary JVLA data (project 13A-26)
for A2390, using the full 1 GHz bandwidth of the L band (1.4 GHz)
\rplus{with 16 spectral windows of 64 channels each}. The observations
were done in 30 minute blocks in each of the B, C and D
configurations. We also calibrated and imaged archival VLA C array
data to check our results for consistency. For A2261, we calibrated
archival VLA data from 4 hours of observations each in the B and D
configuration, with a 100 MHz bandwidth. The D configuration data were
previously analyzed by \cite{2008A&A...484..327V}. We also re-analyzed
archived low-frequency GMRT data, originally published by
\cite{2013A&A...557A..99K}. In \tabl~\ref{tab:sample} we summarize the
\rplus{radio frequency} data used in this Paper.

\begin{table*}
  \caption{VLA, JVLA and GMRT observations used in this Paper.}       
\label{tab:sample}      
\centering                 
\begin{tabular}{l l l r r r r}
\hline\hline  
Object     & project & \rplus{date(s) of}  & facility/array      & no. of    & central  & Bandwidth  \\
           & number & \rplus{observations}  & configuration       & hours     & freq.    &           \\
\hline\hline
A2390 & AF367  & \rplus{28-Apr-2000} & VLA/C  & 1.0 & 1.4 GHz   & 100 MHz   \\   
      & 13A-268 & \rplus{12-Jan-2014; 6-Mar-2015} & JVLA/B & 0.5 & 1.5 GHz   & 1.0 GHz    \\
      & 13A-268 & \rplus{25-Jul-2013; 31-Jul-2013} & JVLA/C & 1.0 & 1.5 GHz   & 1.0 GHz    \\
      & 13A-268 & \rplus{3-Mar-2013; 11-Mar-2013} & JVLA/D & 1.0 & 1.5 GHz  & 1.0 GHz    \\ 
\hline
A2261 & AC696 & \rplus{8-Dec-2003; 3-Jan-2004} & VLA/B & 4.0 & 1.4 GHz  & 100 MHz \\  
      & AC696 & \rplus{20-Aug-2004} & VLA/D & 4.1  & 1.4 GHz  & 100 MHz \\  
      & 16\_117 & \rplus{7-May-2009; 9-May-2009} & GMRT  & 6.0 &  240 MHz  & 6.9 MHz \\
      & 16\_117 & \rplus{7-May-2009; 9-May-2009} & GMRT  & 6.0 &  610 MHz  & 30 MHz  \\
\hline
\end{tabular}
\end{table*}

\rplus{To investigate the dynamical states of our targets, we make
  use of the archival XMM-Newton X-ray observations 0111270101 (for
  A2390) and 0693180901 (for A2261).}

\section{Method}
\label{sec:meth}

\subsection{\rplus{Radio} calibration and imaging}
\label{sec:meth:calimg}

\subsubsection{\rplus{VLA, JVLA}}

Calibration was carried out with standard flux and phase calibrators
using the \casa \rplus{\citep{2007ASPC..376..127M}}  software package.
\rplus{Flux and bandpass calibration at 1.4 GHz were performed against
  the known sources 3C286 (A2261) and 3C48 (A2390), adopting the
  \cite{1977A&A....61...99B} flux scale. Bandpass stability across the
  observations was assumed. Gain phases and amplitudes were calibrated
  every ten to fifteen minutes against bright nearby sources
  (bootstrapped in amplitude against the primary calibrators), and
  were furthermore self-calibrated at the imaging stage, resulting in
  residual phase errors on the order of 5\% for the VLA data.}
\rplus{Excision of radio frequency interference (RFI) was done by
  careful visual inspection of each spectral window, antenna pair and
  correlation. Approximately 40\% of the 1 GHz bandwidth of JVLA was
  completely lost (including the flagging of $5-8$ channels at the
  edges of spectral windows, depending on the bandpass response). In
  total, the excised data did not exceed 60\% for JVLA. For the
  archived VLA data RFI was less of a problem, with approximately 15\%
  of the data being flagged after visual inspection.}

Imaging was carried out in \casa, using the multi-scale
multi-frequency synthesis (MSMFS) CLEAN algorithm
\citep{2010PhDT........73V} to model the spectrum of each clean
component (JVLA) in addition to using multi-scale CLEANing. For JVLA
data, due to the large bandwidth, we used a spectral model, using a
Taylor expansion with two Taylor terms (multi-frequency
synthesis). VLA and GMRT data, due to the limited bandwidth, were
imaged in single frequency mode.

To avoid biases in the images, we avoided creating CLEAN boxes by
visual inspection. Instead, the mask was defined where the flux in the
current image (residual plus clean components convolved with the
restoring beam) was in excess of five times the rms of the residual
image. Cleaning was stopped at the level of the current image noise
estimate, and the mask was re-calculated. We iterated until both the
rms level of the residual and the extent of the mask converged; the
latter in the sense that no new pixels were added to the mask.

We first imaged each data set (by array configuration) separately,
\rplus{using Briggs weighting with \url{robust}=0 and}
performing several cycles of phase-only self-calibration. In the A2390
field, two compact sources residing in the secondary lobes of the
primary beam were modeled and subtracted from the $uv$ plane,
separately for each data set. \rplus{The coordinates of these sources
  are (J2000) [21h52m25.4, +17d34m43] and [21h54m40.2, +17d27m59]. The
  corresponding NVSS (NRAO VLA Sky Survey;
  \citealt{1998AJ....115.1693C}) measurements of integrated flux are
  680.5 mJy and 293.9 mJy, respectively.}

We proceeded to combine the visibilities from all array configurations
and re-imaged the targets. At this point we performed one iteration of
self-calibration on amplitude and phase, with the data averaged over
ten minute chunks.
\rplus{This improved the phase solutions of the visibilities from the
  more extended array configurations, in addition to mitigating
  problems arising from systematic errors in the amplitude calibration
  between the different observations.}

Due to the mismatched bandwidths, we imaged the VLA data separately
from the JVLA data. The integrity of the calibration was checked by
comparing integrated flux densities of compact sources in the images
to those of the NVSS \citep{1998AJ....115.1693C} and FIRST
\citep{1995ApJ...450..559B} surveys.

Relative weights on the visibilities were re-estimated from the data,
after subtraction of a preliminary model of the sky emission, to
improve the rms of the images. For JVLA, this was done from the rms of
the data, per baseline and spectral window in segments of 1 minute of
integration, after subtraction of a source model, thereby improving
the image rms by approximately 20\% on average. For VLA, visibility
weights were computed on a longer time scale (5 minutes) due to the
continuum setup of these observations. This improved the image rms by
approximately 12\% in the A2261 field.

\subsubsection{\rplus{GMRT}} 

\rplus{The GMRT data at 240 and 610 MHz were calibrated and edited
  using the AIPS-based SPAM pipeline as detailed by
  \cite{2016arXiv160304368I}. In short, after inspecting and flagging
  (removing) bad data from dead antennas, corrupt baselines and RFI,
  the primary calibrator 3C48 was used to derive instrumental bandpass
  and complex gain calibrations, adopting the
  \cite{2012MNRAS.423L..30S} flux scale. Several rounds of
  self-calibration, wide-field imaging and flagging of more bad data
  were started off by phase calibrating the target field data against
  a simple sky model derived from other radio surveys (NVSS,
  \citealt{1998AJ....115.1693C}; WENSS, \citealt{1997A&AS..124..259R};
  and VLSSr, \citealt{2014MNRAS.440..327L}). Next, two rounds of
  direction-dependent calibration (peeling), ionospheric modeling
  \citep{2009A&A...501.1185I} and wide-field imaging were
  performed. At this point, all sources beyond a radius of 10 arcmin
  from the primary beam center were removed (e.g.,
  \citealt{2015MNRAS.454.3391B}), thus creating a data set suitable
  for final imaging in CASA.  }


\subsubsection{\rplus{Radio} images}


\rplus{In Table~\ref{tab:imagedetails} we present the properties of
  our most important image products, including all the ones presented
  as figures in this work. We discuss the images further in
  \sect~\ref{sec:res}.} \\

\begin{table*}
    \caption{\rplus{Properties of the VLA, JVLA and GMRT final images presented in this work. The uv data have been tapered to yield the quoted synthesized beams.}}       
  \label{tab:imagedetails}      
  \centering                 
  \begin{tabular}{l r l c r l}
    \hline\hline  
    Object    & central  &  facility/array   & synthesized beam & rms per & Figure \\
    & freq.    &  configuration(s) & FWHM (arcsec)         & beam & \\
    \hline\hline
    A2390 & 1.4 GHz & JVLA / B+C+D  & $30\times30$ &  40 $\mu$Jy & Fig.~\ref{fig:a2390} \\  
    A2390 & 1.4 GHz & VLA / C       & $16\times16$ &  43 $\mu$Jy & (not shown) \\
    \hline
    A2261 & 1.4 GHz & VLA / B+D     & $50\times50$ &  50 $\mu$Jy & Fig.~\ref{fig:a2261} \\
    A2261 & 1.4 GHz & VLA / B       & $15\times15$ &  34 $\mu$Jy & Fig.~\ref{fig:a2261compact} \\
    A2261 & 240 MHz & GMRT          & $15\times15$ &  700 $\mu$Jy & Fig.~\ref{fig:a2261compact} \\
    A2261 & 240 MHz & GMRT          & $28\times28$ &  870 $\mu$Jy & Fig.~\ref{fig:gmrt} \\
    A2261 & 610 MHz & GMRT          & $12\times12$ &  90 $\mu$Jy & Fig.~\ref{fig:gmrt} \\
    A2261 & 610 MHz & GMRT          & $28\times28$ &  160 $\mu$Jy & Fig.~\ref{fig:gmrt} \\
    \hline 
  \end{tabular}
\end{table*}

\subsection{\rplus{Radio} image analysis}
\label{sec:meth:imganal}

In this subsection we describe the sensitive process of
separat\rplus{ing} compact from the extended
emission by means of using different parts of the $uv$ space of the
interferometric observations. Cool core clusters typically feature
bright, central radio sources (e.g., \citealt{1999MNRAS.306..599E};
\citealt{2009A&A...501..835M}), necessitating a very high dynamic
range to separate out the relatively much fainter diffuse
emission. With traditional synthesis imaging, the latter is not
typically attainable with short observations where the $uv$ space is
too sparsely sampled.

\rplus{As it is necessary to not only separate out point-like or very
  moderately resolved sources (such as radio galaxies) but also
  partially extended structures (such as jets and radio lobes) an
  angular scale must be carefully chosen below which the emission is
  considered `compact'. This is difficult to do in general, as radio
  lobes can extend to Mpc scales in extreme cases. In this work, we
  follow the pragmatic approach of inspecting the residual images
  (after compact source subtraction) to visually exclude the
  possibility of residual extended emission from radio lobes
  (\sect~\ref{sec:res}).}

After repeated self-calibration (phase only, \rplus{see
  \sect~\ref{sec:meth:calimg}}) to improve the dynamic range, we used
the long baselines to make models of the compact emission component
(radio galaxies, jets and lobes) at 20 cm in each field, imaging with
Briggs weighting with \url{robust}=0. Given a set of visibilities, the
connection of $uv$ distance with the typical recoverable scale is a
complicated function of the $uv$ coverage and the associated weights
of the visibilities. For example, in an observation containing lots of
radio frequency interference (RFI), predominantly the short spacings
may be either severely decimated by flagging or have low weights due
to low-level residual RFI superficially manifesting itself as a higher
noise level. For these reasons, similar angular to physical scale
conversions still result in dissimilar $uv$ cuts for the different
visibilities. We carefully selected these cuts so as to maximize the
recovery of compact emissions (smaller than a scale of $\sim200$kpc,
which is considerably larger than a radio galaxy but smaller than a
GRH) while at the same time minimizing the contribution from extended
emissions on scales of $\gtrsim0.5$ Mpc in the compact emission image,
resulting in cuts of $>$1.8 k$\Lambda$ for A2390 and $>$1.3 k$\Lambda$
for A2261.  The compact emission models were then Fourier transformed,
de-gridded and subtracted from the visibilities prior to imaging the
diffuse emission using natural weighting.

\rplus{Images of the compact emission were made with the multi-scale
  feature of \casa \url{clean} turned off, building the clean models
  as sums of point sources.}




Each of our fields has a bright central radio source, partially
resolved in the more extended array configurations. 
\rplus{In the A2390 and A2261 fields, respectively, the coordinates of
  these sources are (J2000) 21h53m36.83 +17d41m43.7 and 17h22m17.01
  +32d09m12.9, with the corresponding integrated flux densities (from
  Gaussian fits) 230.6$\pm$1.0 mJy and 18.75$\pm$0.46 mJy at 20 cm.}
Special care was taken to
model these sources, in particular by joint imaging of data from
different array configurations with subsequent self-calibration. We
found that the necessary amplitude adjustments were always less than a
few per cent, yet vital to the accurate modeling of the bright
sources.

Additionally, many types of sources of different spatial extent are
present in the targeted fields. We carried out 
\rplus{three} separate tests for systematic effects in the separation
of the compact emission:

\begin{enumerate}
\item{To test for a possible \textit{loss of flux due to missing short
      spacings}, we inserted the CLEAN model of the extended emission
    into an emission-free direction of the image plane,
    Fourier-transformed this model to the uv plane and made a
    deconvolved image. We repeated this process with slightly
    stretched models of the extended emission, and found no
    significant loss (less than 5\% for both targets, when stretched by
    up to $20\%$).}
\item{To test for \textit{loss of flux due to subtraction of the
      compact emission}, we again Fourier transformed the CLEAN model
    of the extended emission and cleaned using only the long
    baselines, to verify that none of this emission was recovered by
    the process. We found that at most $\sim2\%$ of the diffuse
    emission was recovered in this way. Note, however, that this
    simple test does not account for losses due to substructures in the
    radio emission, that may have been lost prior to modeling this
    emission.}
\item{\rplus{Residual emission from partially subtracted compact
      sources could possibly be hidden in the noise of the residual
      images at some level. We tested the robustness of the compact
      source subtraction by making alternative compact images with a
      (\textit{u,v}) cut corresponding to scales smaller than
      $\sim400$ kpc (rather than the nominal 200 kpc used for our main
      analysis). Imaging the residuals resulted in the extended
      emission being somewhat attenuated, but at a level of less than
      15\% in both targets, suggesting that our choice of 200 kpc is
      robust.}}
\end{enumerate}

We made deconvolved (CLEANed) images at $1.4$ GHz from compact
source-subtracted visibilities, and extracted the diffuse radio flux
by measuring the integrated flux inside an aperture corresponding to
the $2\sigma$-contour of the emission. While this makes our
measurements dependent on signal-to-noise, the difference is less than
$10\%$ when taking a larger aperture.

1.4 GHz radio luminosities\footnote{with this definition we are
  strictly measuring a luminosity density. For simplicity, we stick
  with the term `luminosity' for the remainder of the work.} for the
extended radio emission were computed as
\begin{equation}
  P_{\mathrm{1.4 \, GHz}} \ = \ (4\pi \ D_L^2) \ S_{\mathrm{1.4 \, GHz}} 
  (1+z)^{\alpha-1} ,
\label{eq:lumeq}
\end{equation}
where $S_{\mathrm{1.4 \, GHz}}$ is the integrated flux density, $D_L$
is the cosmological luminosity distance and the factor
  $(1+z)^{\alpha-1}$ is the K-correction, accounting for the observed
flux corresponding to a higher rest frequency. The spectral
  index, $\alpha$, is defined using the convention $S_1 =
  S_0\left(\frac{\nu_1}{\nu_0}\right)^{-\alpha}$.  Uncertainties on
$S_{\mathrm{1.4 \, GHz}}$ and $\alpha$ were propagated through the
calculation. The derivation of spectral indices is discussed in
  the next subsection.

\subsection{Spectral slopes}
\label{sec:meth:alpha}

The spectral index of the extended radio emission in A2390 was
estimated from the wide-band 20 cm data in two ways, which were found
to be consistent.  First, we used the \casa spectral index image
produced by the wide-band imaging algorithm with \url{nterms=2} to
find an average spectral index inside the same aperture used for
extracting the flux, albeit with no reliable uncertainty
estimate. Secondly, we split the data by frequency to make images
centered at 1.25 and 1.81 GHz. The flux in these images were again
measured in the aperture described above, with uncertainties derived
from regions of the images with no apparent signal. With a simple
 \rplus{Monte Carlo approach}, drawing
flux \rplus{values} at the two frequencies from the allowed
ranges, we  \rplus{obtained} a posterior
distribution of the spectral index.
Because residual signals of bright point sources and radio lobes can
\rplus{conceivably} affect the determination of the spectral index, we
also excluded regions around the brightest point sources and
determined the spectral index again. 
\rplus{These regions were constructued as a pixel mask, first choosing
  all pixels exceeding 3$\sigma$ in the image of the compact emission,
  and subsequenly extending the mask to correspond to the resolution
  of the image with the residual (extended) emission}.
This had no effect on the determined values other than a slight
increase in the uncertainty.

As we have only narrow-band data at 1.4 GHz for A2261, the spectral
slope was estimated in conjunction with the GMRT data, from which we
do not have a significant detection of the extended component from
either frequency. We measured the signal in the GMRT images in two
ways: First, we subtracted the compact emission using the clean
components of images made using a lower uv cut of 0.5k$\Lambda$. To
avoid giving to much weight to the longest baselines (sensitive only
to the most compact emission), we tapered the visibilities at
1.4k$\Lambda$ for these images of the compact emission. We subtracted
these components from the visibilities and made new images, now
tapered at 0.8k$\Lambda$ (to recover any extended emission), and
proceeded to directly measure the flux of the possible extended
emission, in the same aperture as for the 20cm image. To rule out
contamination due to residual emission from the complex structure to
the North-West of the cluster center \rplus{(see
  Fig.~\ref{fig:a2261compact})}, we additionally made use of a second
method in which we measured the flux in an aperture excluding this
region. \rplus{Specifially, the region corresponds to an excess of
  3$\sigma$ or more in any of the three images of the compact
  emission}. The flux was then corrected upwards to match the flux
ratio of the VLA image in the corresponding regions. The measured
values (see Table~\ref{tab:results}) using the two methods were found
to be consistent.

We tested how well the GMRT data can recover a structure of the size
and morphology found at 20cm by injecting the 20cm model, scaled with
a spectral index $\alpha=1.2$, and measuring the difference in the
resulting images. We found relatively high flux recovery ratios: 0.85
at 610 MHz and 0.89 at 240 GHz. The measured signals were corrected
upward by the inverses of these factors.

Finally, to determine the mean spectral slope of the A2261 extended
emission given the flux measurements at the three different
frequencies, we made use of a Monte Carlo Markov chain to fit for the
spectral slope (in conjunction with a flux normalization which was
subsequently marginalized over).

\subsection{\rplus{X-ray data analysis}}
\label{sec:meth:x}

\rplus{The latest versions of the Observation Data Files were
  downloaded from the XMM-Newton Science Archive
  (XSA\footnote{http://nxsa.esac.esa.int/nxsa-web/}) and reduced with
  version 14.0.0 of the XMM Science Analysis System
  (XMM-SAS\footnote{http://www.cosmos.esa.int/web/xmm-newton/sas}). Calibrated
  event-lists were produced for the EPIC cameras using the standard
  processing script \textit{emchain}/\textit{epchain}, based on the
  up-to-date calibration database as of January 2015.  The event-lists
  were filtered for time intervals affected by particle contamination
  using the method described by \cite{2016A&A...592A...2P}.
  For A2261, this resulted in a loss of $\sim$8\% of the EPIC pn
  data. For A2390, about 50\% of the data were rejected for all three
  instruments. The usable (effective area weighted) exposure time was
  9.3ks for A2390 and 25.1ks for A2261. We used the test developed by
  \cite{2008A&A...478..575K} to identify EPIC-MOS CCDs in the
  so-called anomalous state, which shows enhanced soft band emission,
  and cross-checked the results via visual inspection of images in the
  200-900 eV band. This resulted in the exclusion of CCD4 from the
  MOS2 camera in the observation of A2261.}

\rplus{For the morphological analysis (\sect~\ref{sec:morph}), we
  used the [0.4-1.25] keV band. 
  The automated source detection pipeline of
  \cite{2006MNRAS.372..578P} was used to create preliminary source
  masks, which were modified by visual inspection to appropriately
  retain possible substructures in the ICM. The masked areas were
  replaced by the values of randomly selected pixels in surrounding
  annuli. 
  We modeled the background using calibration observations taken with
  the filter wheel in \texttt{Closed} position. First, we reprojected
  the stacked event-lists to the same average attitude and applied a
  rescaling factor for each CCD based on the data recorded in the
  unexposed CCD corners in the same imaging band\footnote{For the
    EPIC-PN camera, one single rescaling factor was computed and
    applied to each quadrant. For the EPIC-MOS central CCDs, which
    have no unexposed corner, we used the average rescaling factor of
    the outer CCDs whose noise properties best correlate with them
    according to \cite{2008A&A...478..575K} - (2, 3, 6, 7) for MOS1
    and (3, 4, 6) for MOS2.}. In a second step, we used the outer
  parts of each pointing to estimate average sky backgrounds at the
  locations of the clusters and added them to the instrumental
  background maps. While sky levels estimated in this way may be
  biased high by residual emission in the cluster outskirts, any such
  bias would be negligible compared to the cluster emission within
  500\,kpc which we analyze in \sect~\ref{sec:morph}. Finally, we
  produced background substracted and exposure corrected surface
  brightness images, applying a light ($\sigma=3^{\prime\prime}$)
  gaussian smoothing.}

\section{Results}
\label{sec:res}

\begin{figure*}
  \hspace{-0pt}
\includegraphics[width=\columnwidth, clip=true, trim=-70 0 0 0]{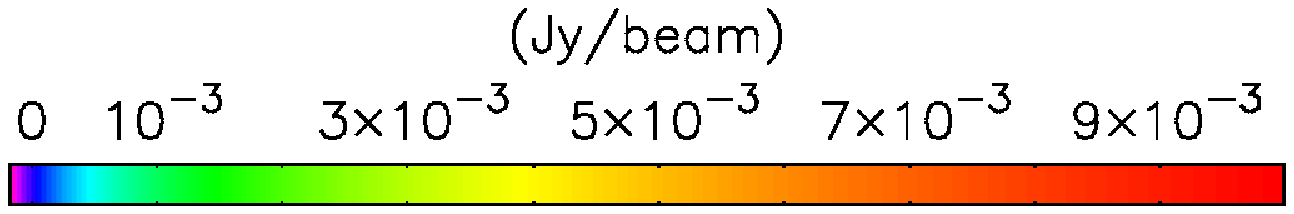} \\
\vspace{-0pt}
  \includegraphics[width=\columnwidth, clip=true, trim=0 300 100 50]{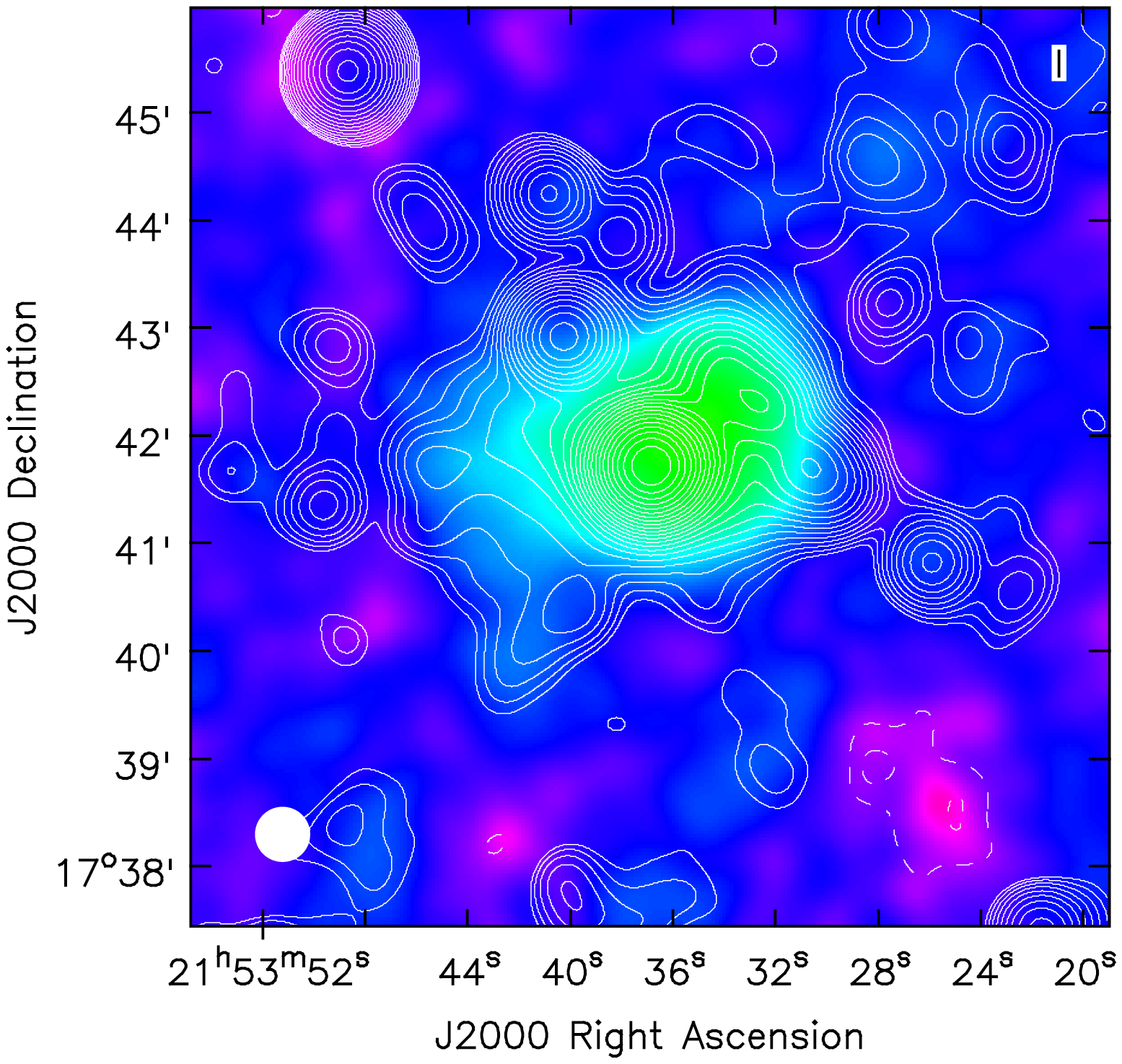}
  \includegraphics[width=\columnwidth, clip=true, trim=0 300 100 50]{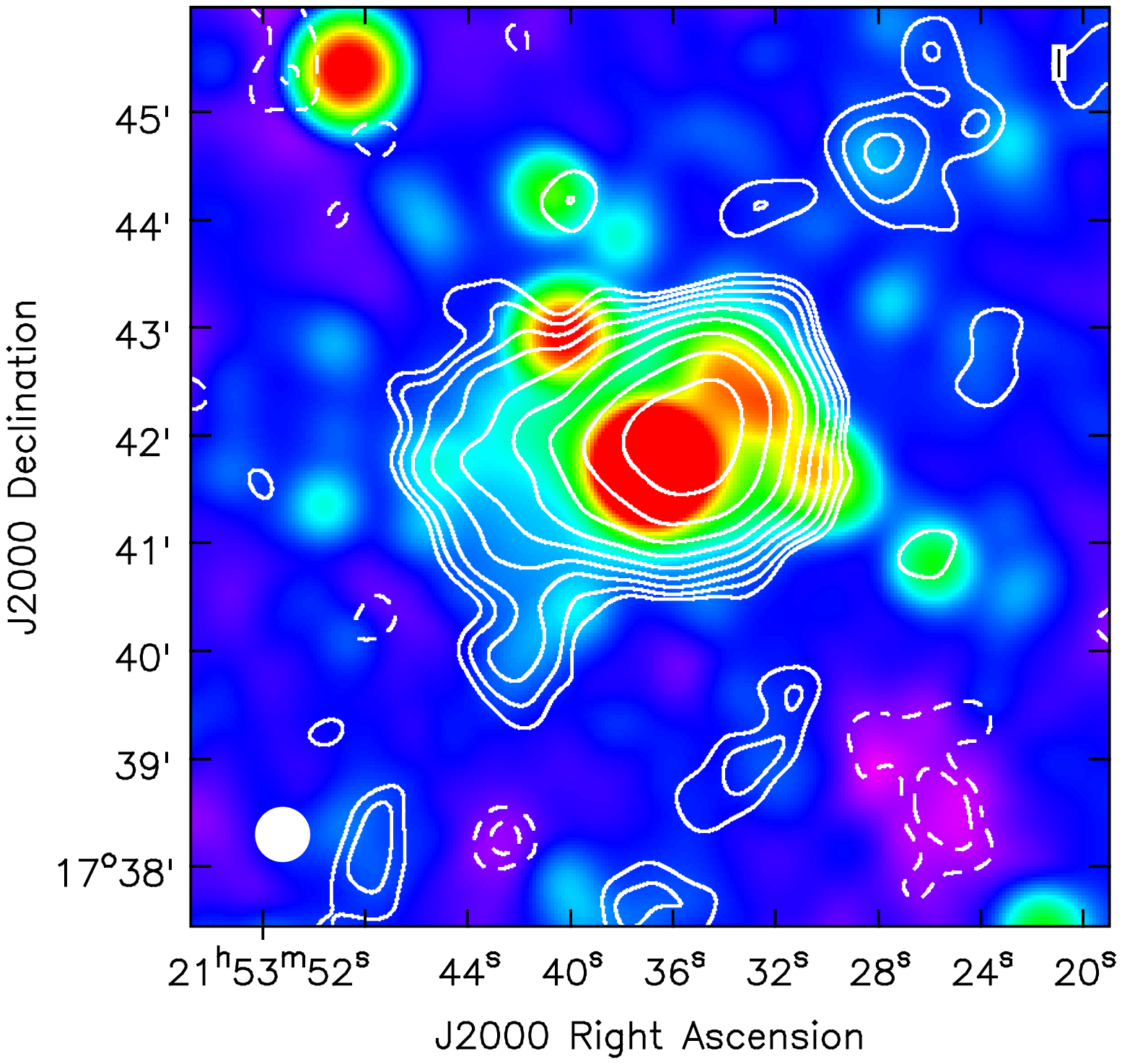}
  \caption{Deconvolved JVLA radio images of A2390 at 20 cm. The full
    image was produced from all baselines using Briggs weighting with
    \protect\url{robust}=1 (\casa convention), and is shown as
    contours in the left panel and as colors in the right panel. The
    diffuse extended emission was imaged after subtracting the
    compact emission (see main text) and is shown as colors in the
    left panel and as contours in the right panel. The images have
    been smoothed to the same resolution, and both have an rms noise
    level of 40 $\mu$Jy per $30^{\prime\prime}\times30^{\prime\prime}$
    restoring beam (FWHM). Contours are drawn at $\pm \text{rms}
    \times \sqrt{2}^n$, with $n=\{2,3,4,...\}$. The (non-linear) color
    scale is the same in both panels.}
  \label{fig:a2390}
\end{figure*}

\begin{figure*}
  \hspace{-0pt}
\includegraphics[width=\columnwidth, clip=true, trim=-70 0 10 0]{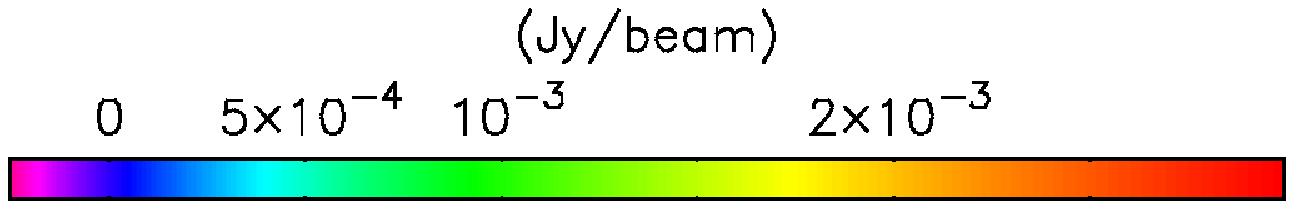} \\
\vspace{-0pt}
  \includegraphics[width=\columnwidth, clip=true, trim=0 300 100 50]{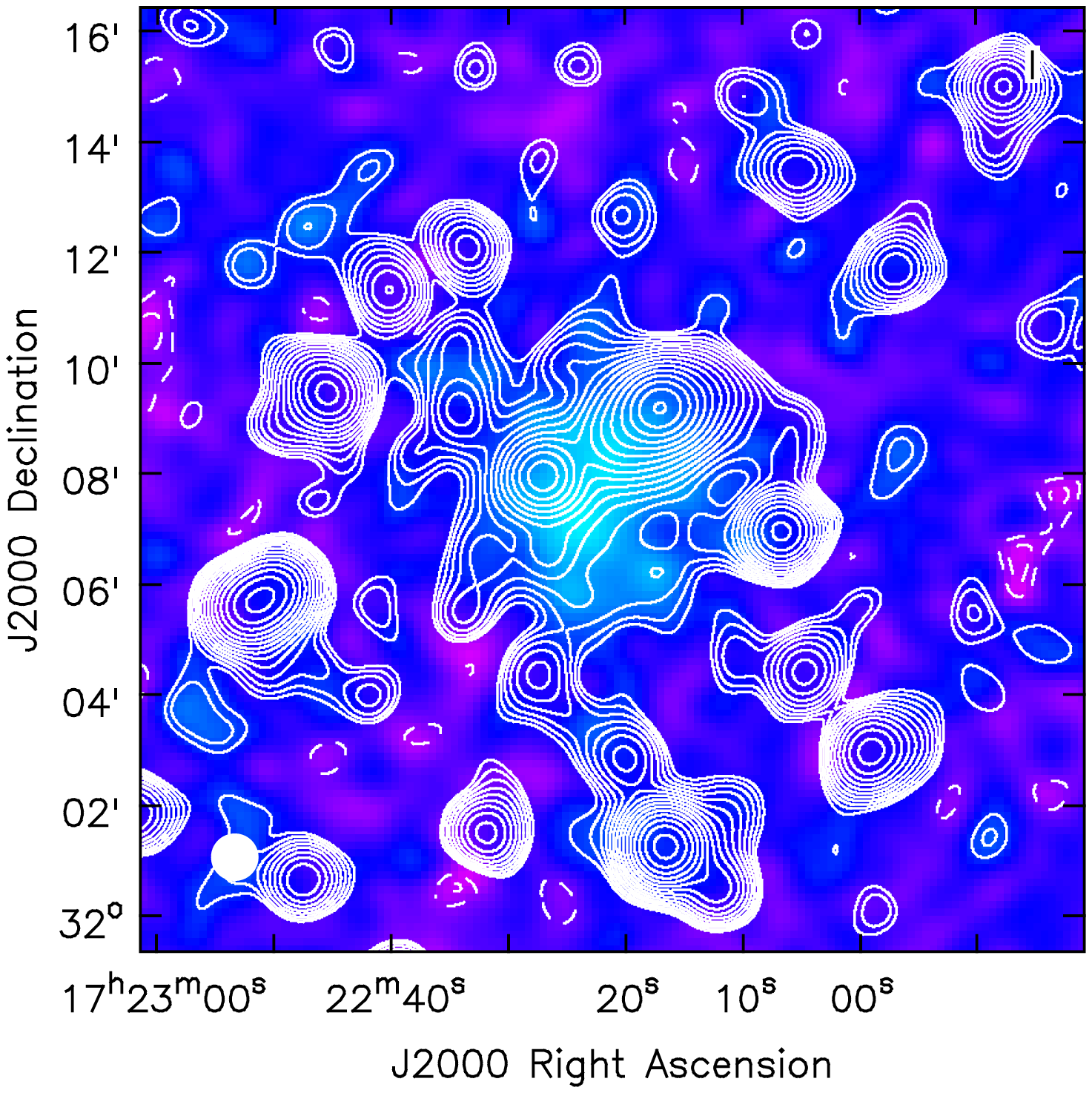}
  \includegraphics[width=\columnwidth, clip=true, trim=0 300 100 50]{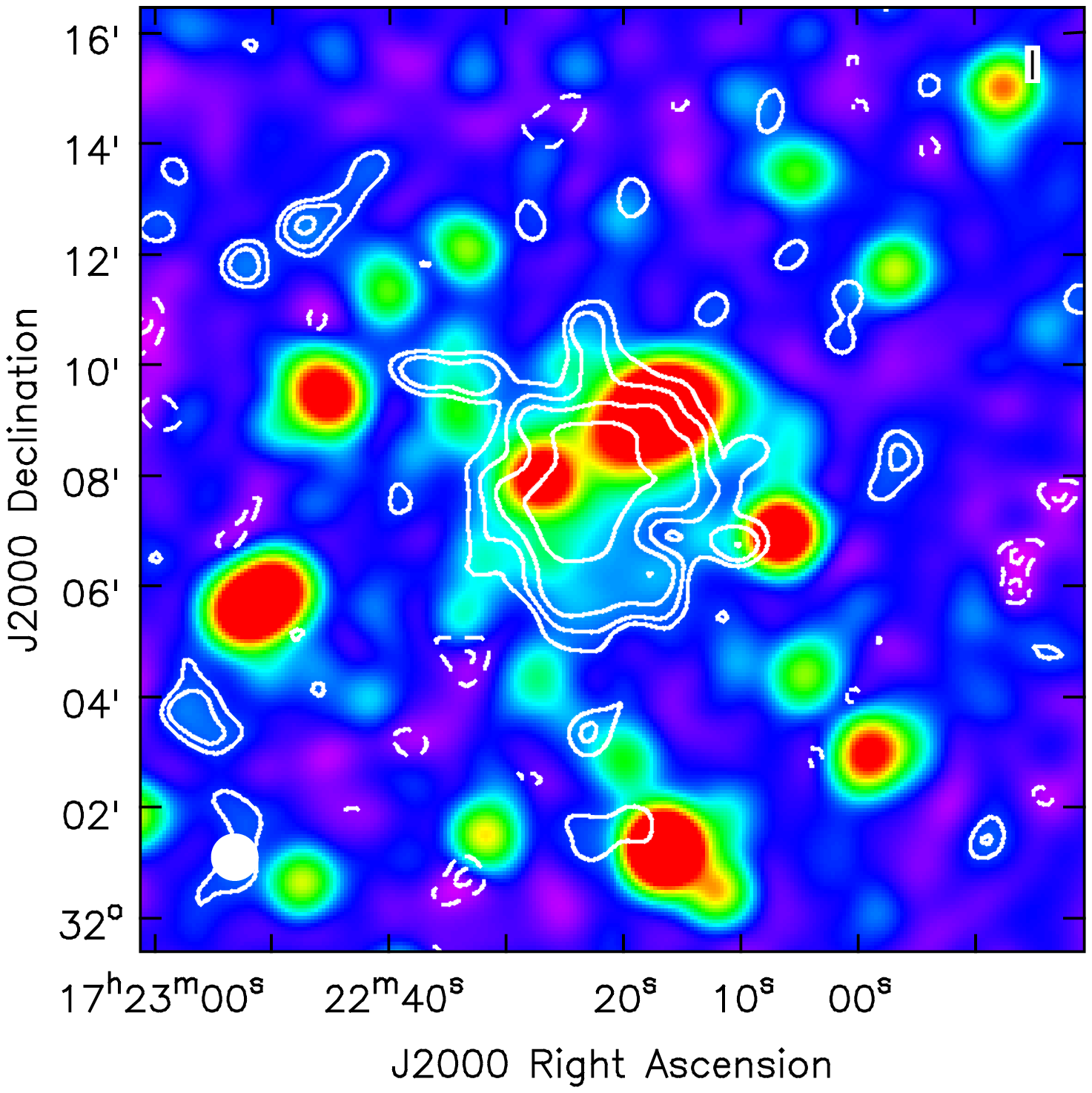}
  \caption{\rplus{Deconvolved JVLA radio images of A2261 at 20 cm. The
      full image was produced from all baselines using Briggs
      weighting with \protect\url{robust}=1 (\casa convention), and is
      shown as contours in the left panel and as colors in the right
      panel. The diffuse extended emission was imaged after
      subtracting the compact emission (see main text) and is shown as
      colors in the left panel and as contours in the right panel. The
      images have been smoothed to the same resolution, and both have
      an rms noise level of 50 $\mu$Jy per
      $50^{\prime\prime}\times50^{\prime\prime}$ restoring beam
      (FWHM). Contours are drawn as in Fig.~\ref{fig:a2390}. The
      (non-linear) color scale is the same in both panels.}
    }
  \label{fig:a2261}
\end{figure*}

\begin{figure}
  \includegraphics[width=\columnwidth, clip=true, trim=30 350 100 0]{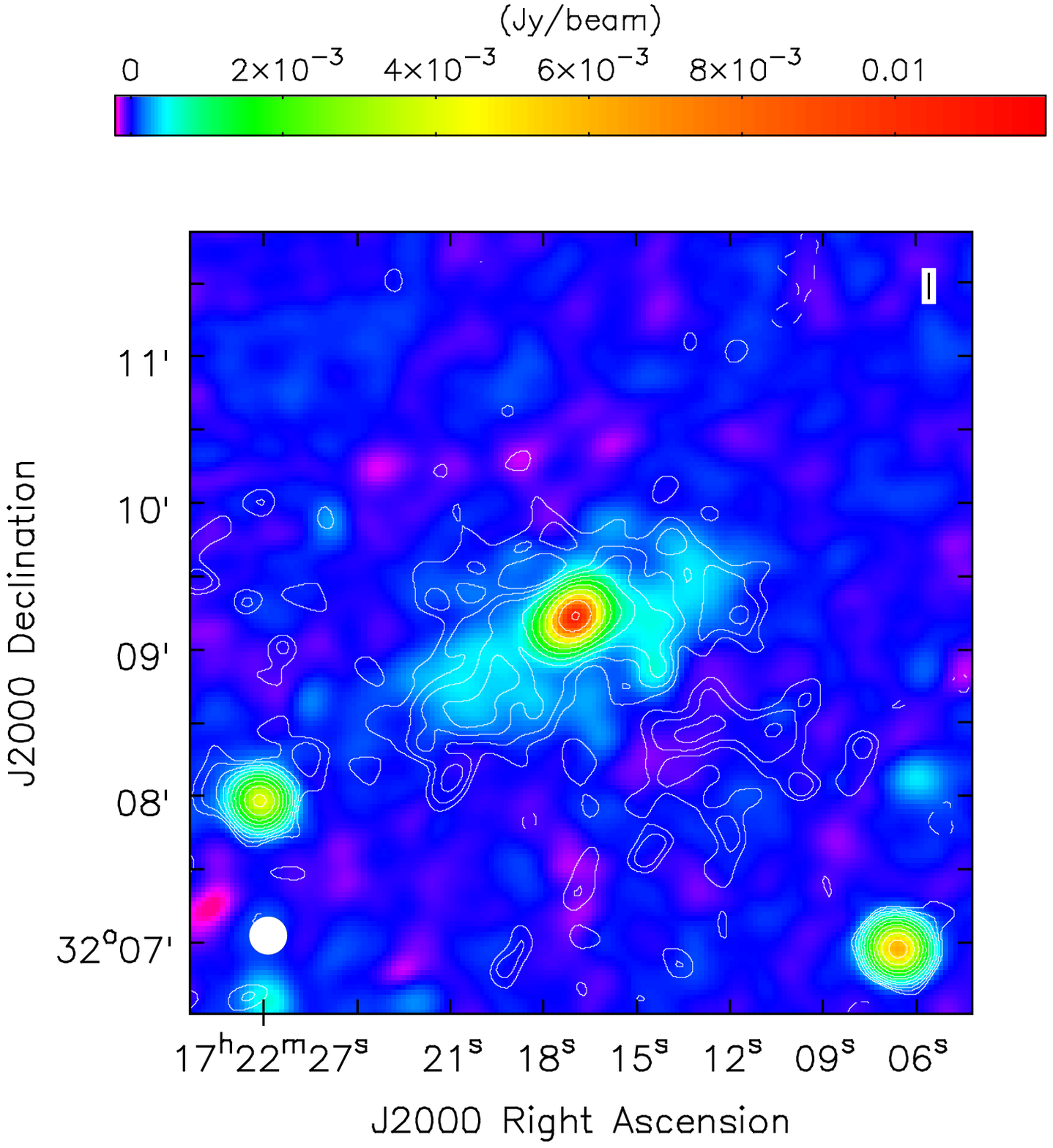}
  \caption{Higher resolution image of the central region of A2261,
    \rplus{highlighting the bright compact emission of and around the
      bright central source}. \rplus{The} background image \rplus{was
      made} from the VLA B  \rplus{configuration} data
    \rplus{at 1.4 GHz} (rms 34 $\mu$Jy per beam), \rplus{and the}
    foreground contours \rplus{from the} GMRT \rplus{data} at 240 MHz
    (rms 0.7 mJy/beam). \rplus{Contours are drawn at $\pm \text{rms}
      \times \sqrt{2}^n$, with $n=\{2,3,4,...\}$.}  Both images have a
    common resolution of $15^{\prime\prime}\times15^{\prime\prime}$
    (FWHM).}
  \label{fig:a2261compact}
\end{figure}

In this section we present radio images of the targets and the
associated physical parameters derived from the images. VLA L-band
images of A2390 and A2261 are shown in Figures \ref{fig:a2390} and
\ref{fig:a2261} (array configurations are listed in Table
\ref{tab:sample}).  The image of A2390 was smoothed to highlight the
$\sim$1 Mpc scale emission. For A2261 we show a heavily tapered and
smoothed image in Figure \ref{fig:a2261} to bring out the relatively
featureless emission on a scale of several arcminutes and contrast it
to other features, including the structure in the central region. The
latter is highlighted in Figure \ref{fig:a2261compact}, where it is
shown in higher resolution in conjunction with a GMRT image at 240
MHz. 
The emission extended at the Mpc scale is clearly distinct from the
more compact components, judging in particular by the relative lack of
features in the former and the efficiency with which marginally
resolved features outside the central regions of the targets are
removed by the subtraction of compact components. The morphology of
the emission in the images made from the visibilities with the compact
emission removed resemble typical GRHs from the literature in both
size and shape. \rplus{Notably, the compact structure around the
  bright, central source does not align with the peak of the extended
  emission, suggesting that this is indeed a separate emission
  component.}

\begin{figure*}
\includegraphics[width=\columnwidth, clip=true, trim=30 260 100 0]{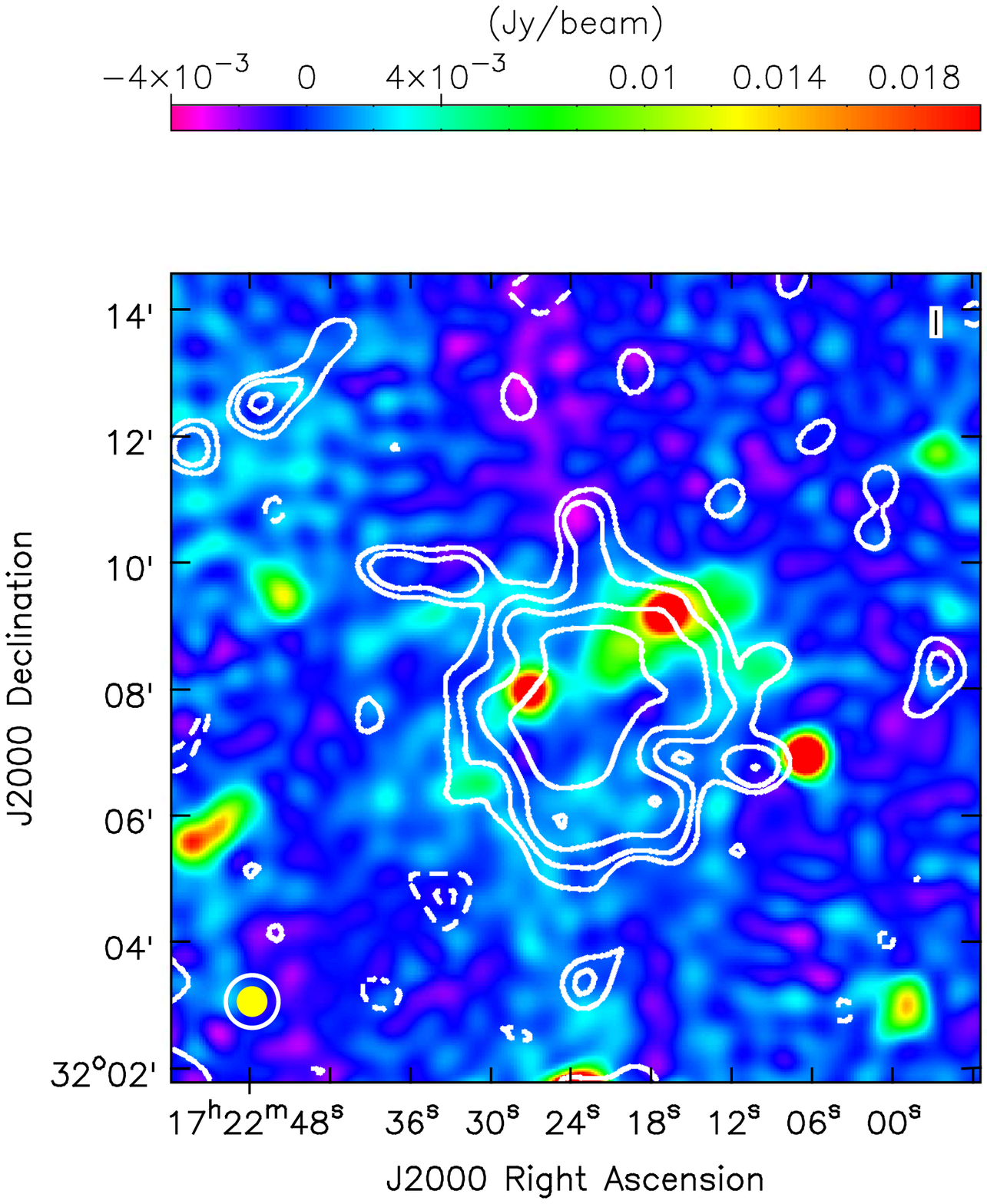} 
\includegraphics[width=\columnwidth, clip=true, trim=30 260 100 0]{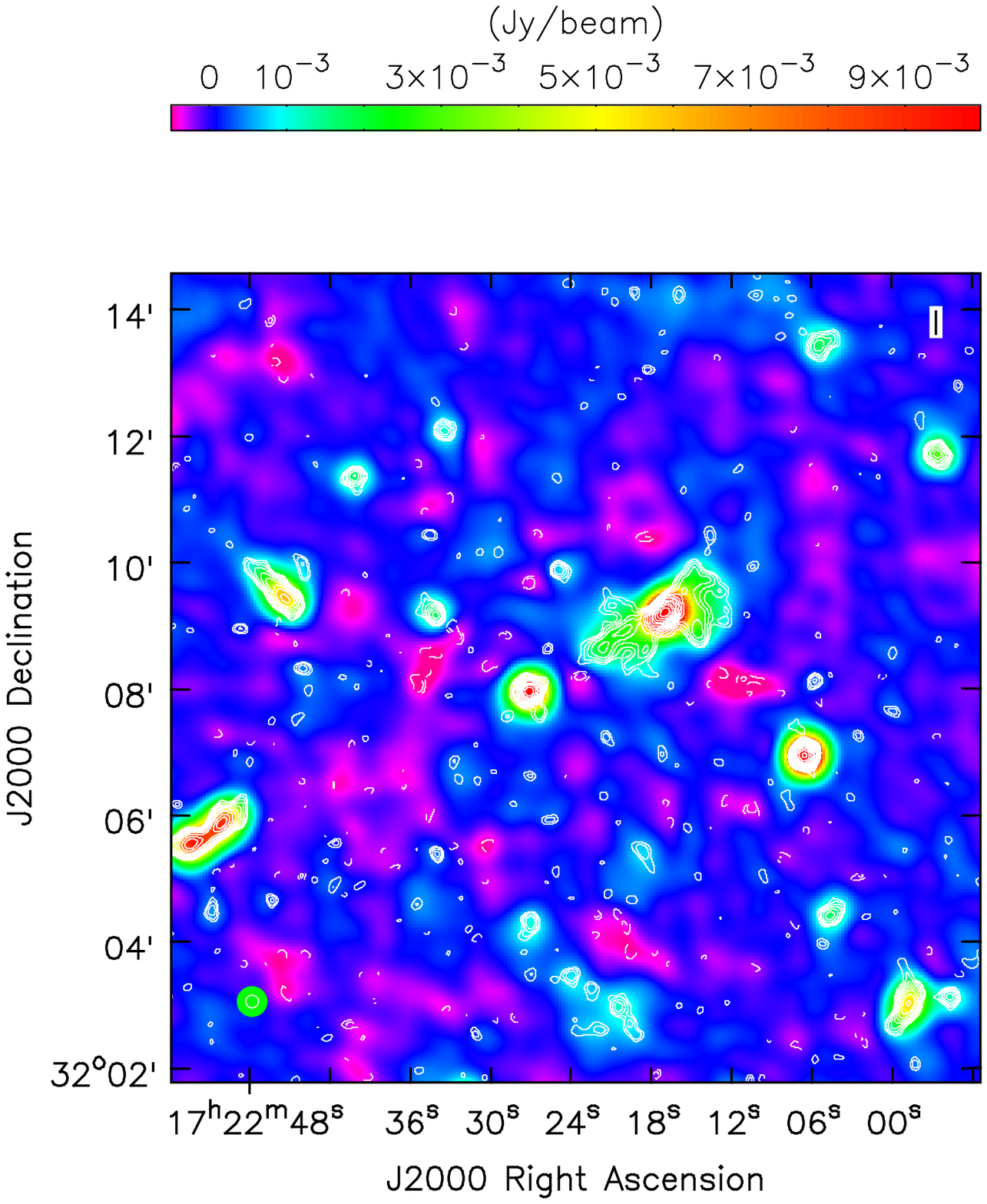}
\caption{GMRT images (colors) of the A2261 field at 240 and 610 MHz,
  using natural weights. Both images were smoothed to a common
  resolution of $28^{\prime\prime}\times28^{\prime\prime}$ (FWHM). The
  240 MHz image (\textit{left}) has a noise rms of 870 $\mu$Jy per
  beam, while the rms level in the 610 MHz image is 160 $\mu$Jy per
  beam. The contours in the left image represent the VLA image of the
  extended emission at 1.4 GHz
  ($50^{\prime\prime}\times50^{\prime\prime}$ FWHM). The contours
  overlaid on the 610 MHz image in the right panel correspond to the
  native (not convolved) resolution of this image, with an rms noise
  level of 90 $\mu$Jy per $12^{\prime\prime}\times12^{\prime\prime}$
  (FWHM) beam. Filled black beam ellipses correspond to color images,
  white beam ellipses to contours.}
  \label{fig:gmrt}
\end{figure*}

GMRT images of A2261 at 240 and 610 GHz are shown in
Fig.~\ref{fig:gmrt}. Although an extended emission component is not
apparent in these images, this is expected due to the relatively high
level of noise: in spite of a GRH being expected to be brighter at
these frequencies, the signal-to-noise ratio is in fact lower. The
latter becomes obvious from Fig.~\ref{fig:spectrum}, where we show the
resulting spectral fit for the A2261 halo. We note here also that the
flux at 610 GHz is consistent with the $2\sigma$ upper limit of
\citet{2013A&A...557A..99K}.

The relevant physical parameters derived from our analysis are
summarized in Table~\ref{tab:results}. The largest linear scale (LLS)
of each target was measured from the $2\sigma$-contours of the diffuse
emission component images, adjusting for the synthesized beam. For
convenience we also give the exact conversions of angular and physical
scales given the concordance cosmology. The spectral slope of A2390 is
derived from the 20 cm JVLA data alone, as we have no lower frequency
data.

\begin{table*}
  \caption{Assumed and derived physical parameters of our target clusters. LLS: largest linear scale. Flux densities were integrated over an aperture corresponding to a significance of 2 sigma.}       
\label{tab:results}      
\centering                 
\begin{tabular}{l r r r r r r r r r r@{}l}
  \hline\hline  
  Object & Redshift & $D_{A}$ & Mpc/ & LLS & LLS & $S_{240\,\text{MHz}}$ & $S_{610\,\text{MHz}}$ & $S_{1.4\,\text{GHz}}$ & $P_{1.4\text{GHz}}$  &  Spectral~ & slope  \\
         & $z$      & [Mpc] & arcmin & [$^{\prime}$]  & [Mpc]  & [mJy] & [mJy] & [mJy] &  [$10^{23}$~W/Hz] &  $\alpha$  &      \\
  \hline\hline
  A2390  & \rplus{0.228}    & \rplus{743} & \rplus{0.216} & 3.6 & \rplus{0.8} & & & 16.80$\pm$0.37  & 30.71$\pm$1.29    &  $1.60$ & $\pm0.17$ $^{(1)}$ \\   
  A2261  & 0.224    & 733 & 0.213 & 5.7 & 1.2 & 79$\pm$34 & 6.6$\pm$4.6 & 4.37$\pm$0.35   & 7.04$\pm$0.80     &  $1.20$  & $^{+0.23}_{-0.50}$ $^{(2)}$\\
  \hline
\multicolumn{11}{l}{\rplus{$^{(1)} 1-2$ GHz in-band spectral index}} \\
\multicolumn{11}{l}{\rplus{$^{(2)} 240$ MHz$-1.4$ GHz spectral index}} \\
\end{tabular}
\\
\end{table*}

\section{Discussion}
\label{sec:discussion}
   
\subsection{Comparison with previous results}

\subsubsection{A2390}

The A2390 VLA data at 20 cm were previously analyzed by
\cite{2003A&A...400..465B}, who identified the radio emission as a
mini-halo with an integrated signal of 63$\pm$3 mJy. Because this
value is inconsistent with our result, we re-analyzed the VLA data
used by Bacchi et al. After careful flagging and self-calibration, we
were able to deconvolve an image consistent with our deeper JVLA
result. While we confirm the existence of a ``hole'' in the emission
north of the BCG, we find no evidence of filaments extending to the
north as reported by Bacchi et al. The large inconsistency could be
due to a different definition of the compact emission component. This
is difficult to verify as Bacchi et al. do not show an image for which
the compact emission has been subtracted; nor do they specify the
weighting scheme used in the imaging or the uv tapering used to
separate out the emission from compact sources.

The largest physical scale of the extended emission found by Bacchi et
al. is 550 kpc, which differs from our estimate of 800 kpc. This
corresponds directly to the reported difference in the angular size of
the radio emission (after accounting for the respective cosmological
models): Bacchi et al. found the largest extent to be $2^{\prime}$,
compared to our measurement of $3.6^{\prime}$. This is likely
explained by the lack of short spacings in the earlier analysis.

\subsubsection{A2261}
\label{sec:a2261}

The VLA data of A2261 were previously analyzed by
\cite{2008A&A...484..327V}, who found hints of extended emission but
concluded that further analysis would be necessary to confirm the
result. We have independently confirmed the presence of the extended
emission by modeling the compact component from VLA B and D
configuration data. Our  \rplus{tests for the
  robustness of the separation of compact and diffuse emission
  components}, the methods of which are outlined in
section~\ref{sec:meth:imganal}, indicate \rplus{that the present
  detection of an extended emission component is robust}.

\begin{figure}
\centering
\includegraphics[width=\columnwidth]{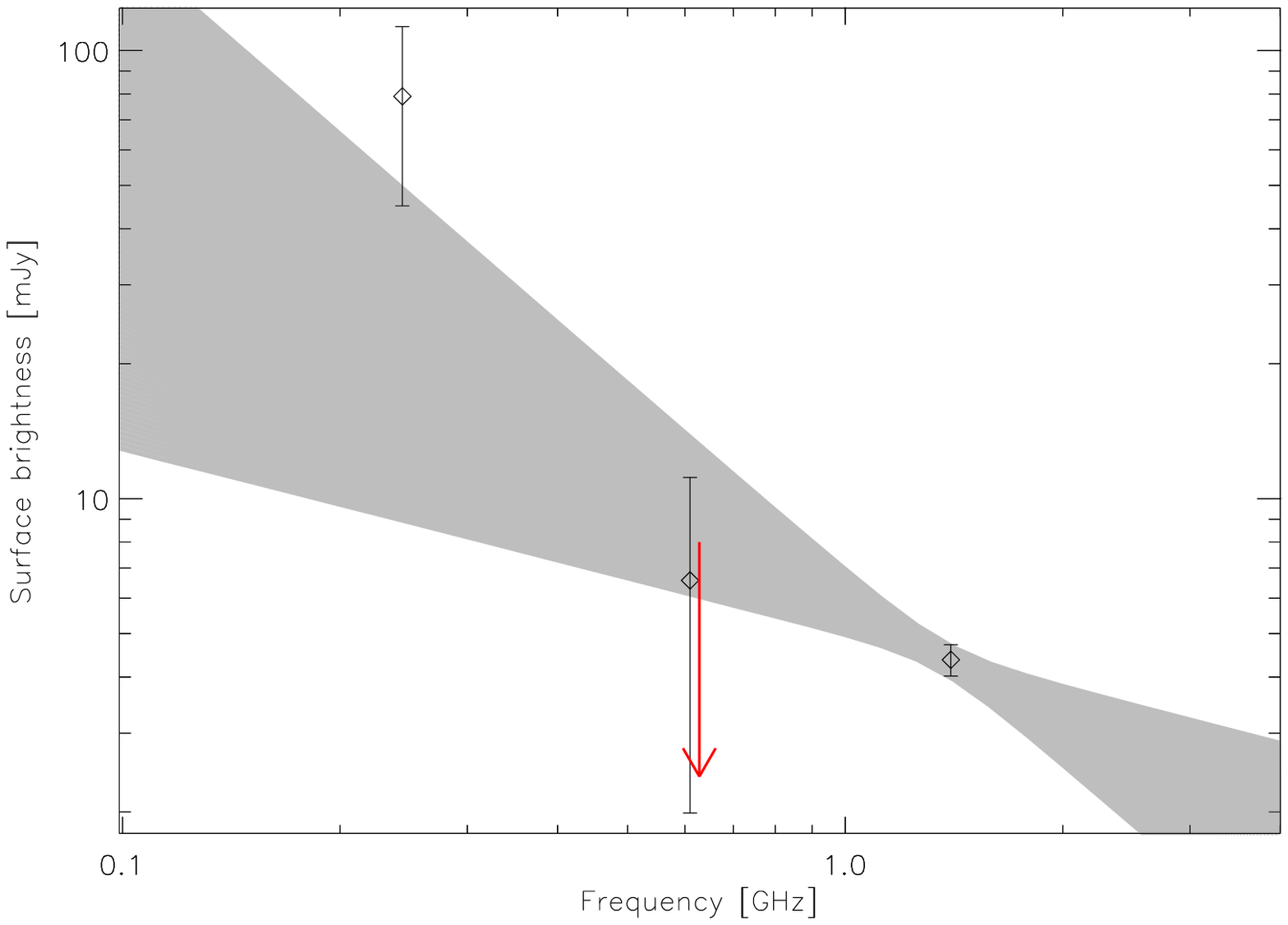}
\caption{Flux measurements (error bars show uncertainties at one
  sigma) of the diffuse radio emission in A2261 based on archival GMRT
  and VLA data, and the resulting spectral fit. The gray shaded region
  show the $1\sigma$ credibility range for the best-fit power-law. The
  large uncertainties from the GMRT data allow us to place only weak
  constraints on the spectral slope. The previous non-detection by
  \citet{2013A&A...557A..99K} at 240 MHz, marked by the red arrow for
  a $2\sigma$ upper limit (shifted slightly in frequency for clarity),
  is consistent with our results.}
\label{fig:spectrum}
\end{figure}

The GMRT data at 240 MHz were first studied by
\cite{2013A&A...557A..99K}, who reported non-detections of radio halos
from the 240 and 610 MHz observations. In particular, upper limits of
8 mJy and 6 mJy at 610 MHz were derived from injection of radio halos
with linear sizes of 1.22 Mpc and 1.0 Mpc, respectively. No upper
limits were specified for the 240 MHz band. Our measurements are
marginally consistent with the results of Kale et al., although our
610 MHz flux measurement is lower than expected from our best-fit
spectral slope. A much shallower spectral slope, consistent with the
results of Kale et al., is allowed by our analysis
(Fig. \ref{fig:spectrum}).

\subsection{Scaling properties}

To shed further light upon whether the extended emission found in our
targets should be classified as giant radio halos, we investigate
whether the measured luminosities are in the expected range for radio
halos, given their linear projected sizes as well as the X-ray
luminosities and integrated Comptonizations of the host clusters.

Linear radii $R_H$ are computed following the method of
\citet{2007MNRAS.378.1565C}, using the geometric mean of the largest
and smallest radii at the $3\sigma$ level of the image to characterize
the size of a radio halo.  Integrated Comptonizations within $r_{500}$
are adopted from the \textit{Planck} PSZ catalog
\citep{2014A&A...571A..29P}, where we use the \textit{Planck}
  X-ray derived values of $r_{500}$. We compute the
\textit{intrinsic} Comptonization
\begin{equation}
\label{eq:yszdef}
Y_{\text{SZ}} = E(z)^{-2/3} D^2_A Y_{{500}}, 
\end{equation}
where $D_A$ is the angular diameter distance and $E^2(z) =
\Omega_M(1+z)^3 + \Omega_{\Lambda} + \Omega_k(1+z)^2$. 
We obtained X-ray soft band (0.1$-$2.4 keV) luminosities $L_X$ for
A2390 and A2261 from the BCS \citep{1998MNRAS.301..881E} and eBCS
\citep{2000MNRAS.318..333E} clusters catalogs.



\begin{figure*}
\centering
\includegraphics[width=18cm, clip=true, trim=0 0 0 0]{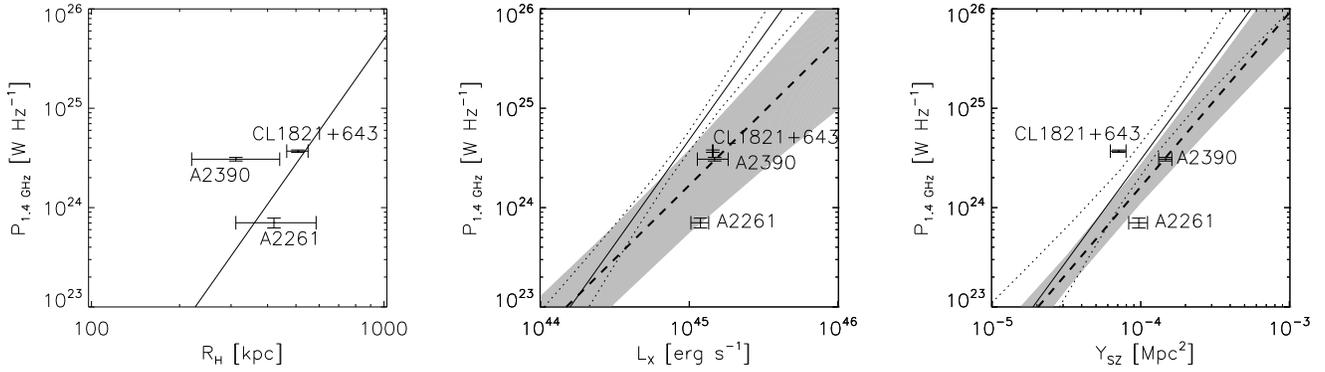}
\caption{Comparison of the radio luminosity of our targets with known
  scaling relations in terms of size of the halo $R_{H}$, X-ray
  luminosity $L_{X}$, and integrated Comptonization $Y_{\text{SZ}}$.
  For comparison, we also include the system CL1821+643 from
  \citet{2014MNRAS.444L..44B}. \textit{Left:} The solid line
  represents the best-fit relation of
  \citet{2007MNRAS.378.1565C}. \textit{Middle and right}: The thin
  solid lines represent the best-fit relations of
  \citet{2013ApJ...777..141C}, with confidence intervals demarcated by
  dotted lines. Also shown are the best-fit relations of
  \citet{2014MNRAS.437.2163S}, as thick dashed lines (uncertainties as
  shaded regions). The Sommer \& Basu relations are more consistent
  with our current measurements, possibly because these were derived
  from a mass-limited sample of clusters. }
  \label{fig:scaling}
\end{figure*}

\rplus{In Fig.~\ref{fig:scaling} we show the scaling of radio
  luminosity with X-ray luminosity and integrated Comptonization for
  our two targets. We also include the data for the earlier
  identification of a GRH in the cool-core cluster CL1821+643
  \citep{2014MNRAS.444L..44B}. We compare our measurements to the
  scaling relations derived by \cite{2014MNRAS.437.2163S} and
  \cite{2013ApJ...777..141C}. The measured scaling relations have
  strong intrinsic scatter in the radio luminosity, which can be
  either physical in origin or due to the difficulty in isolating and
  quantifying the full extended emission using interferometers, which
  in turn can lead to an underestimation of the systematic errors in
  flux measurements. Given the scatter levels reported by
  \cite{2014MNRAS.437.2163S} and \cite{2013ApJ...777..141C}, the
  presently reported radio luminositied are consistent with the
  scaling relations of those works. While a clear scaling of the radio
  power with mass proxies has been established for GRHs, a
  corresponding property has not been demonstrated for
  radio mini-halos.}


Comparing to the scaling relations of \cite{2013ApJ...777..141C},
A2390 and A2261 appear radio under-luminous. \rplus{Apart from the
  aforementioned large scatter and the fact that we have only three
  data points in this case,} these scaling relations were derived from
a fit explicitly excluding the radio non-detections, in a sample not
complete within a mass selection. It is thus possible that the Cassano
et al. scaling relations are biased towards radio luminous
objects. While \cite{2013ApJ...777..141C} also derived relations for
cool-core corrected X-ray luminosities, with similar results, such an
analysis does not remove the effect of the selection.

\subsection{Dynamical state}

We discovered what appear to be giant radio halos in two cool-core
clusters of galaxies. Taken together with the results of
\cite{2014MNRAS.444L..44B}, there are three known such cases. Most
clusters hosting a cool core are known to be dynamically relaxed
\citep{1992MNRAS.258..177E,2001MNRAS.322..589A,2005MNRAS.359.1481B,2005ApJ...628..655V},
however, the connection between the dynamical state of a cluster and
the presence of a cool-core is not straightforward.  In this section
we examine the empirical evidence available for assessing the
dynamical state of these two clusters. To this end we analyze archival
XMM-Newton data to compute the values of standard X-ray morphological
indicators as well as azimuthal variations in the X-ray brightness.

\subsubsection{Previous X-ray and optical analysis} 

\cite{2000MNRAS.315..269A} identified cool cores in both A2261 and
A2390 from {\it Chandra} data. The measured temperature and entropy
profiles can also be seen from the Archive of Chandra Cluster Entropy
Profiles (ACCEPT\footnote{http://www.pa.msu.edu/astro/MC2/accept/};
\citealt{2009ApJS..182...12C}), showing clear central temperature drops
in both targets. In this work, we quote entropies measured at $20$
kpc radii from \cite{2013ApJ...767..116M}.

A2390 has a low central entropy of $31.6\pm1.1$ keV
cm$^2$. \cite{2001MNRAS.324..877A} determined a cooling radius of
$175^{+40}_{-6}$ kpc, at which the cooling time first exceeds the
Hubble time. \cite{2008MNRAS.385..757D} determined the cooling radius
to be $60.91 \pm 1.16$, defined where the cooling time is 3
Gyr. Regardless of the definition of the cooling radius, the radio
emission detected in the present work is significantly larger than the
radius of cooling and AGN feedback.  Although classifying A2390 as a
relaxed cluster, \cite{2008MNRAS.383..879A} note\rplus{d} substructure
within certain position angles from the \textit{Chandra}
images. \cite{2001MNRAS.324..877A} found substructure in the surface
brightness on scales $\gtrsim 2^{\prime}$, possibly suggesting a
locally disturbed or not fully relaxed dynamical
state. \rplus{\cite{2013A&A...549A..19W} classified A2390 as an
  `intermediate' cluster based on a morphology estimator based on the
  peak of the P3/P0 profile computes in different apertures (see
  \sect~\ref{sec:morph} for an explanation of the power ratio P3/P0).}
A stellar population analysis of the central BCG of A2390
\citep{2008MNRAS.389.1637B, 2016MNRAS.456.1565L} revealed mostly young
($\sim100$ Myr) stars, \rplus{leading to its classification} as a
blue-core system with a short cooling timescale.

The cool-core in A2261 is less prominent than the one in A2390, with a
central entropy of $60.0 \pm 9$ keV cm$^2$. \cite{2008MNRAS.389.1637B}
classif\rplus{ied} this as a red-core system based on the optical
spectra of its central BCG, meaning one that is not actively forming
stars due to the cooling of its gas. Nevertheless, it shows a regular,
round morphology (within the central $\sim$1 Mpc), and based on the
small level of substructures and a small centroid shift of the BCG
from the X-ray peak ($\sim 0.4$ kpc; \citealt{2008MNRAS.389.1637B}),
it has been classified as a relaxed system
\citep[e.g.,][]{2008ApJS..174..117M}.

\subsubsection{X-ray morphological estimators}
\label{sec:morph}

In hierarchically formed structures in the universe, substructures
will be present at some level in any system. We therefore choose to
rely largely on quantitative measures of relaxation and
disturbance. \cite{2010ApJ...721L..82C} showed that galaxy clusters
hosting radio halos can be differentiated with respect to their
dynamical state characterized using different methods. Following this
approach, we computed the power ratio $P_3/P_0$ from a multi-pole
decomposition of the projected mass distribution
\rplus{\citep{1995ApJ...452..522B}} inside an aperture of radius 500
kpc; the centroid shift $w$ from the standard deviation of centroids
computed from a series of apertures out to a radius of 500 kpc; and
the concentration parameter $c$, defined as the ratio of the peak
surface brightness (inside a radius of 100 kpc) and the ambient
surface brightness (inside a radius of 500 kpc). \rplus{The choice
  of 500 kpc apertures makes a direct comparison with
  \cite{2010ApJ...721L..82C} possible.}

\rplus{To compute the concentration parameter $c$, we estimated the peak of
the cluster emission iteratively by smoothing the surface brightness
map to increasingly higher resolution. The flux is simply integrated
around this peak within apertures of 100 and 500kpc as described by
\cite{2010ApJ...721L..82C}, and $c$ is defined as the ratio of the two
estimates.}

\rplus{For the $w$ and the $P_3/P_0$ measurements, we followed the
  suggestion of \cite{2010A&A...514A..32B} to fix the center of the
  aperture at the centroid of the emission obtained by minimizing the
  dipole moment $P_1$. We defined the centroid shift as the rms of the
  distance between the X-ray peak and the flux weighted centroid while
  the aperture size was varied \citep{2006MNRAS.373..881P}.  With our
  XMM-Newton data we only estimated this rms from 10 sub-apertures
  \citep{2013A&A...549A..19W}, i.e. between 50 and 500 kpc with
  increments of 50kpc.  The power ratios were obtained form the
  surface brightness $S(x)$ over the whole aperture
  $R_{ap}$=500\,kpc.}

\rplus{Uncertainties on all parameters were estimated via Monte
  Carlo simulations. For $c$ and $w$, we obtained a photon model of
  the whole observation by multiplying the smoothed surface brightness
  image with the exposure map and adding our background model.  A
  thousand Poisson realisations of the model were generated and
  analyzed in the same way as the true observation. For the power
  ratio the process is a bit different as the bias due to Poisson
  noise must also be corrected for \citep{2013A&A...549A..19W}. Rather than
  performing new simulations, we randomized the azimuthal angle of the
  photons, permitting to estimate and substract the average power
  ratio due to shot noise for a perfectly symmetric structure in
  addition to estimating the uncertainty.}

\rplus{The power ratio and the centroid shift estimators are, by their
  definition, not sensitive to the presence of substructures occuring
  exactly along the line of sight. Merger events can thus not be ruled
  out categorically by these estimators. The concentration parameter,
  converesly, is sensitive also to line-of-sight occurences, and is
  used here as a measure of whether the gas in the cluster core has
  been disrupted by a recent merger event (low gas
  concentration). }

\rplus{Some mergers have been found to yield inconclusive results
  using morphological estimators from X-ray photometry (e.g.,
  \citealt{2016MNRAS.458.3083S}). In such cases, a spectroscopic
  survey of the cluster member galaxies can yield more accurate
  information on the dynamic state
  \citep[e.g.][]{2008A&A...479....1J,2009ApJ...693..901O}.}

\begin{figure}
\centering
\includegraphics[width=\columnwidth, clip=true, trim=20 0 120 0]{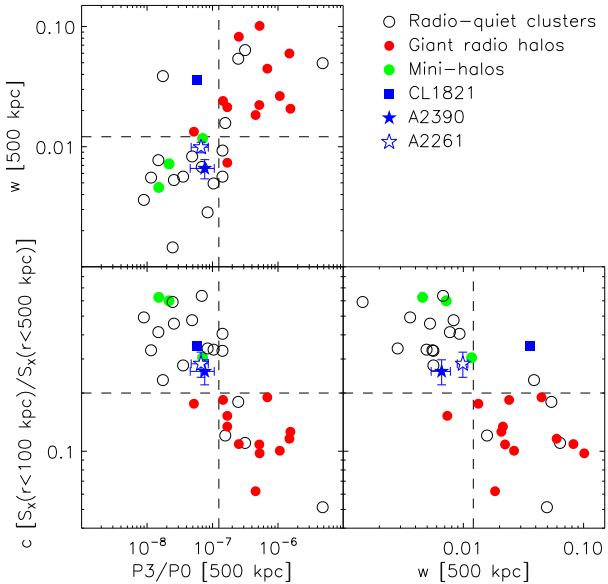}
\caption{Morphological estimators: power ratio $P_3/P_0$, centroid
  shift $w$ and concentration parameter $c$. The data points have been
  adopted from \citet{2010ApJ...721L..82C}, without error bars for
  clarity. Our own data have been added (error bars indicate
  statistical and systematic uncertainties added in quadrature), as
  well as the data on the cool-core cluster CL1821+643 from
  \citet{2014MNRAS.444L..44B}. From this analysis, the targets
  considered in this work have more relaxed morphologies than typical
  for clusters hosting GRHs. }
  \label{fig:morph}
\end{figure}

Because the resolution can have an effect on the parameters,
especially the power ratio, we estimate systematic uncertainties by
further downgrading the resolution. In Figure~\ref{fig:morph} the
computed morphological estimators are shown with error bars indicating
systematic uncertainties added in quadrature to the statistical
uncertainties.

Typically, ``radio loud'' clusters have high power ratios and centroid
shifts combined with low gas concentrations; all characteristic for
dynamically disturbed systems and empirically evident from the data of
\cite{2010ApJ...721L..82C}, adapted in Figure~\ref{fig:morph}. A2261
and A2390 reside in the central region of the joint parameter space,
between known halo clusters and known ``radio quiet'' clusters. Both
targets have small centroid shifts, characteristic for dynamically
relaxed systems and uncharacteristic for radio loud clusters. The gas
concentration parameter $c$ likewise suggests a relaxed morphology,
although this is somewhat inconclusive because this measure is
severely affected by the XMM resolution in the $100-500$ kpc apertures
chosen by \cite{2010ApJ...721L..82C}. To overcome this problem we also
compute concentrations in the larger apertures of $(0.15,1)\times
r_{500}$, where we find $c=0.40 \pm 0.003 \pm 0.031$ for A2390 and
$c=0.43 \pm 0.002 \pm 0.027$ for A2261 (statistical followed by
systematic uncertainties due to the resolution). Comparing with the
analysis of \cite{2013ApJ...777..141C}, these values place both
targets at the boundary of typical radio halo clusters and typical
``radio quiet'' clusters.

The power ratio $P_3/P_0$, while placing A2390 and A2261 closer to
radio-quiet clusters, is also severely affected by the XMM
resolution. However, simulations indicate that this quantity
fluctuates rapidly during mergers \citep{2011MNRAS.418.2467H} and may
not be a reliable tracer for an ongoing merging activity.

\subsubsection{X-ray brightness images}
\label{sec:xbright}

In addition to the `global' morphological indicators of clusters,
X-ray brightness images can reveal important clues of ongoing merger
activities, like bow shocks and plumes behind a merging sub-structure,
or ripple-like features from gas sloshing at the cluster core.  We
create `residual images' of X-ray surface brightness, after
subtracting out the main cluster component, to look for such features.

To create X-ray brightness residual maps, we first fit double
elliptical $\beta$-models to the X-ray photometry in the 0.5$-$2 keV
band. The fit was carried out by making a count rate model which was
tested against the data using the Cash C statistic. The high
signal-to-noise allowed fitting the slope, core radius, normalization,
eccentricity, inclination angle and centroid of each elliptical
$\beta$ model, in addition to a constant sky background. The
instrument background was also taken into account.  We subtracted the
count rate models from the raw images and applied exposure correction.

\begin{figure*}
        \centering
        \hspace{-0.0cm}
        \begin{subfigure}[b]{0.5\textwidth}
                \includegraphics[width=\textwidth, clip=false, trim=0 -14 26 0]{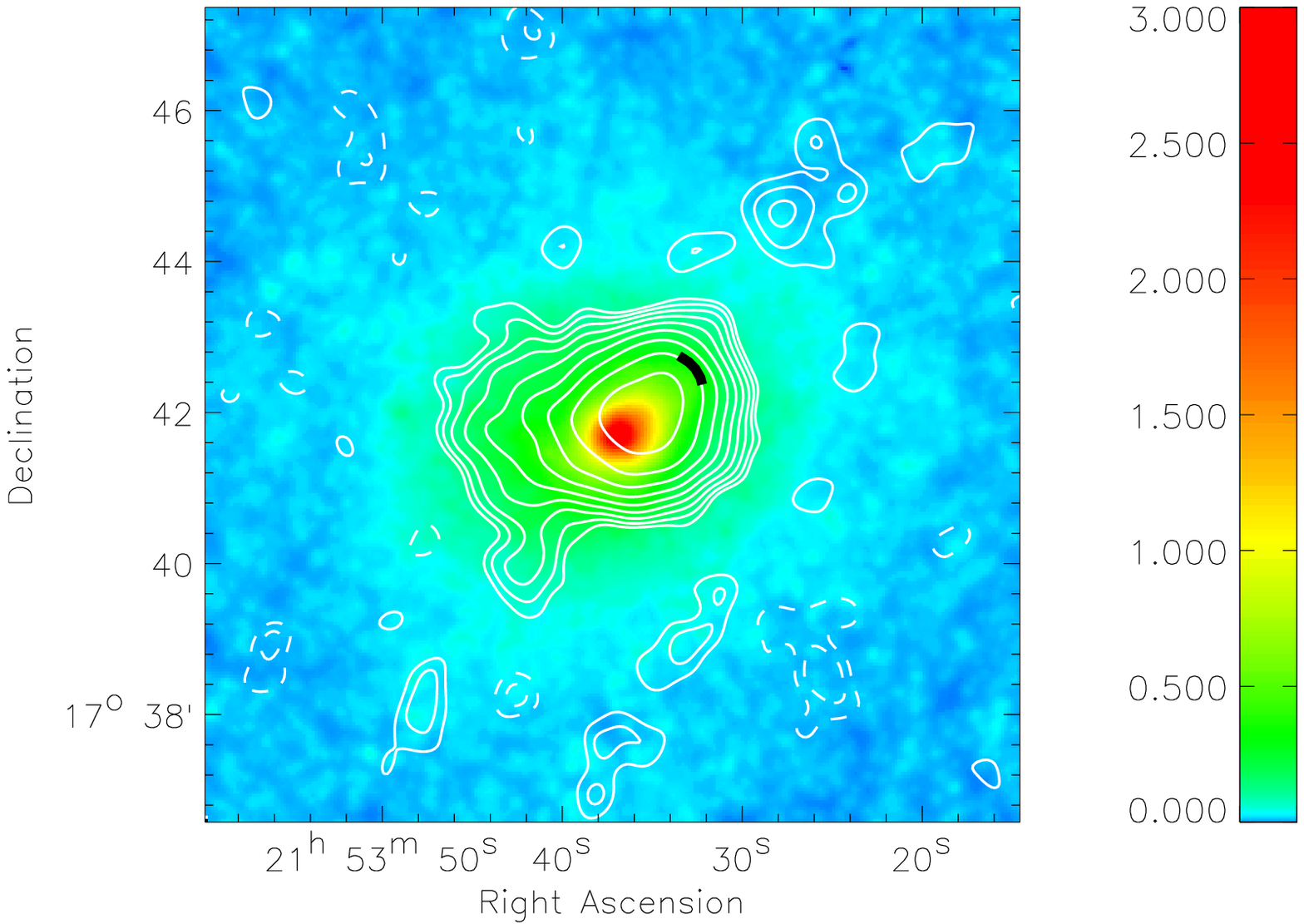}
                \caption{}
                \label{fig:xray:2390:full}
        \end{subfigure}%
        \begin{subfigure}[b]{0.5\textwidth}
                \includegraphics[width=\textwidth, clip=false, trim=0 -14 26 0]{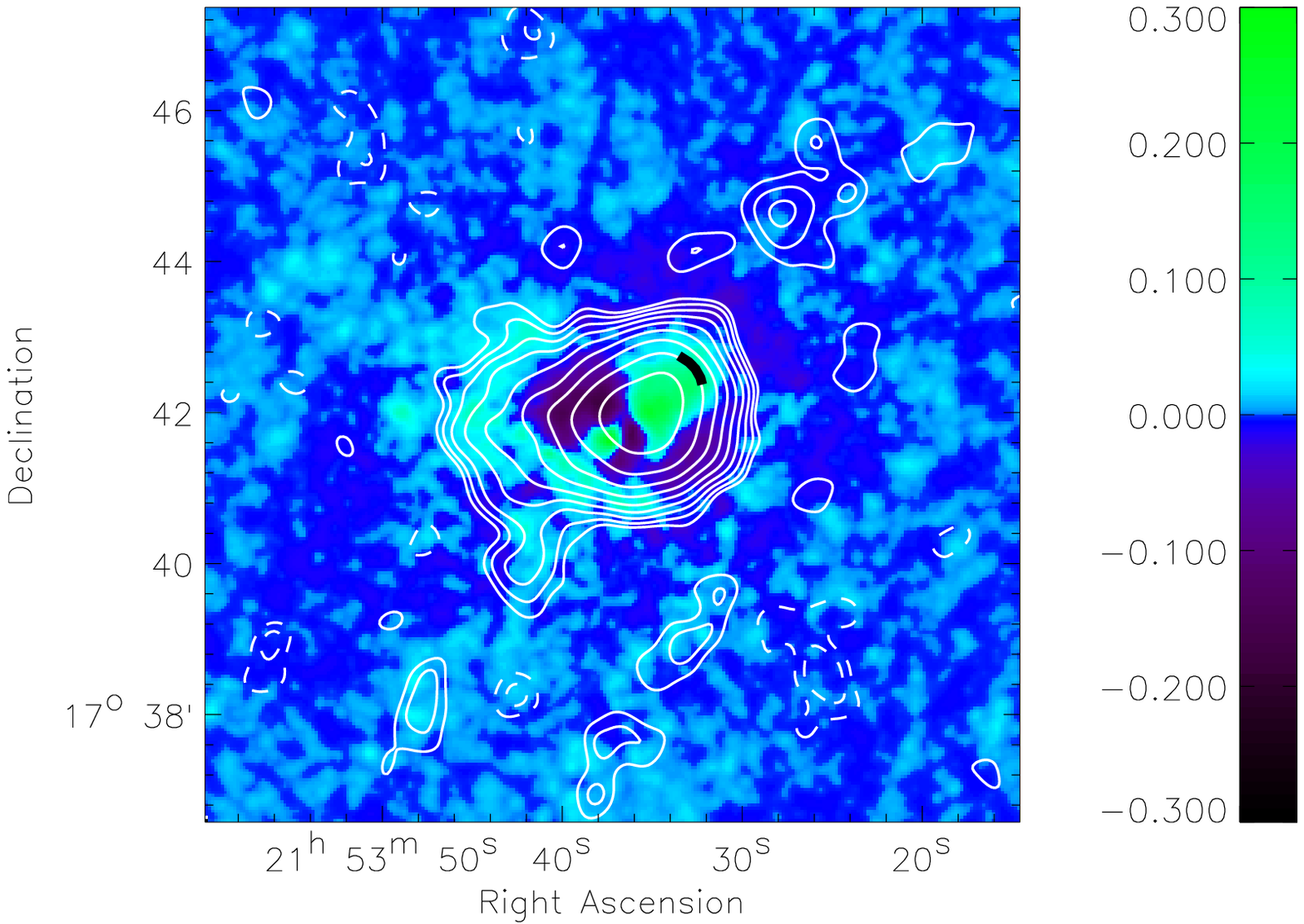}
                \caption{}
                \label{fig:xray:2390:res}
        \end{subfigure}
        \\
        \hspace{-0.0cm}
        \begin{subfigure}[b]{0.5\textwidth}
                \includegraphics[width=\textwidth, clip=false, trim=0 -14 26 0]{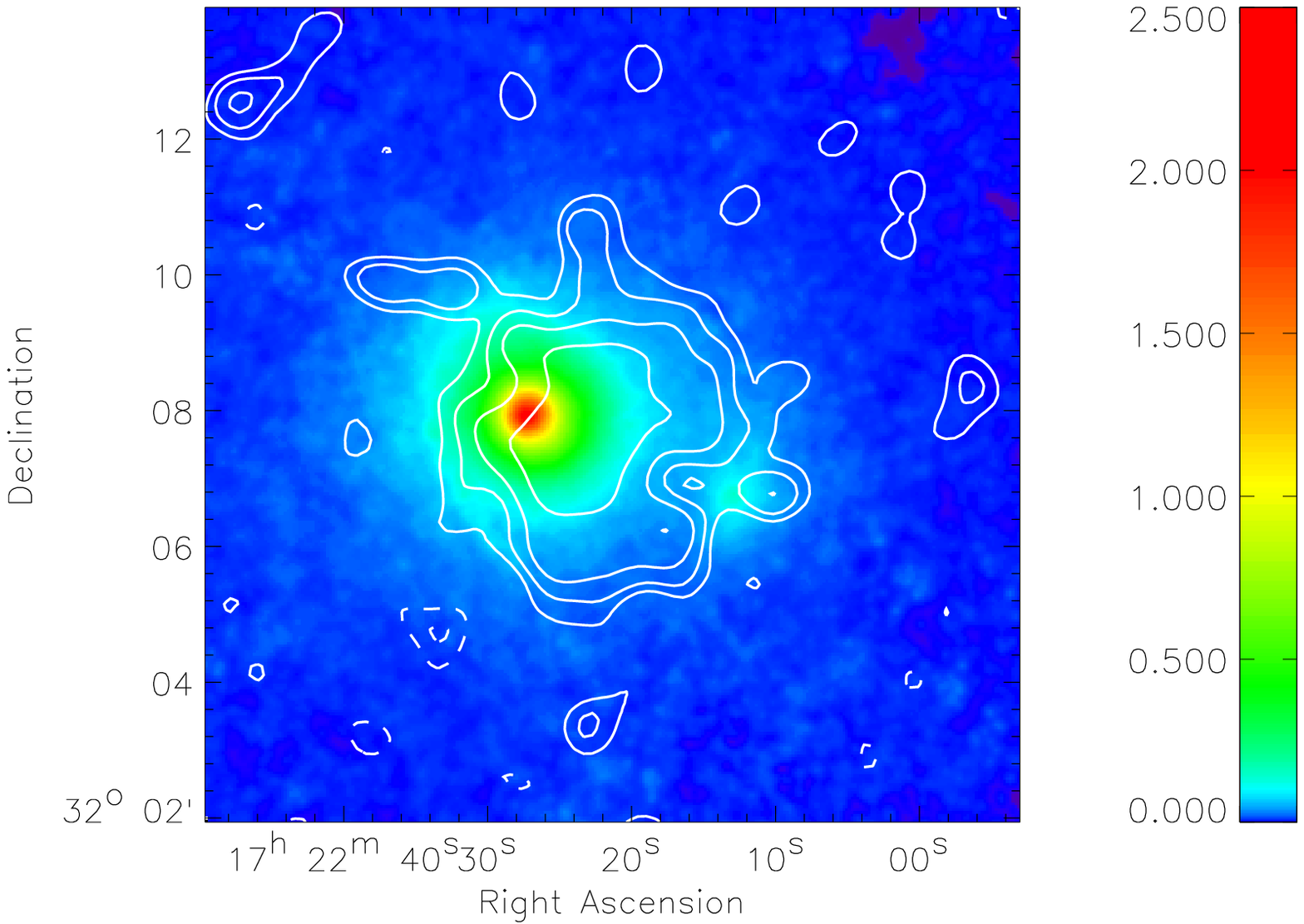}
                \caption{}
                \label{fig:xray:2261:full}
        \end{subfigure}%
        \begin{subfigure}[b]{0.5\textwidth}
                \includegraphics[width=\textwidth, clip=false, trim=0 -14 26 0]{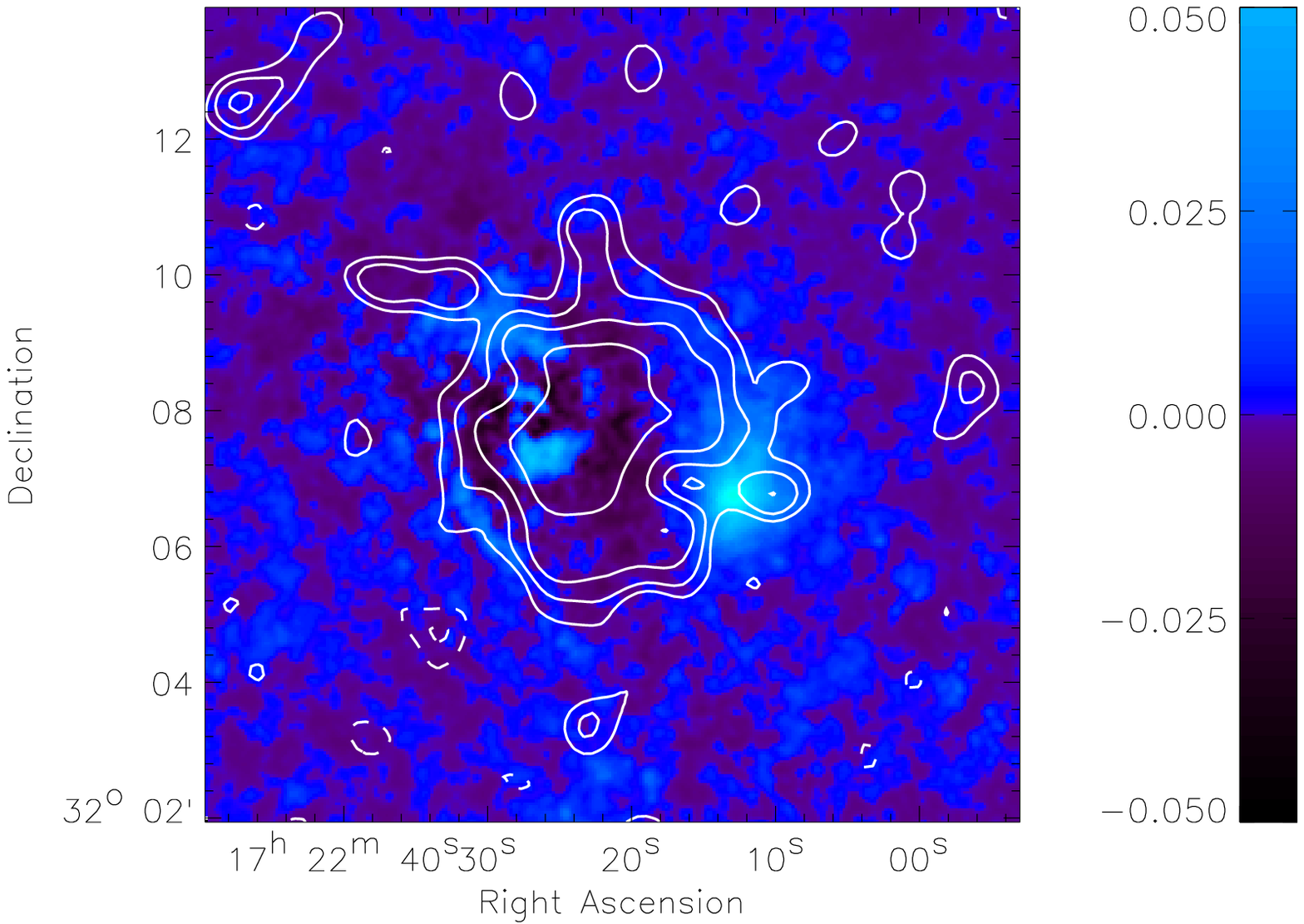}
                \caption{}
                \label{fig:xray:2261:res}
        \end{subfigure}
        \caption{Exposure-corrected XMM count rate images (left) and
          residual images after subtraction of a double $\beta$ model
          (right) of A2390 (top) and A2261 (bottom). For each target
          the color scale is the same in the left and right images,
          with arbitrary units of X-ray surface brightness. We
          indicate the extended radio emission with contours. For
          A2390 we also indicate the position of the sharp brightness
          discontinuity discovered by \citet{2015Ap&SS.359...61S}.}
        \label{fig:xray}
\end{figure*}

The residual images are shown in Fig.~\ref{fig:xray}. In the case of
A2390, the double $\beta$ model does not seem to provide a sufficient
fit to the data, in agreement with the findings of
\cite{2001MNRAS.324..877A}. Nevertheless, the amplitude of the
residuals are small $-$ the relative brightness of the brightest
positive residual signal (north-west of the cluster center) to the
full X-ray emission is less than 1.5\%, leading to maximum
contribution of this substructure to the gas mass of less then 2\%
given the scaling of gas mass with X-ray luminosity
\citep[e.g.][]{2011A&A...526A.105Z}. We do not see any obvious
sub-cluster or bow-shock like features that might indicate an ongoing
merger. The region around the detected radio emission also does not
reveal signs of sharp discontinuities that can indicate sloshing,
although it is possible that the XMM-Newton resolution is not
sufficient for detecting such sharp features.  Indeed,
\cite{2015Ap&SS.359...61S} identified one such feature from the {\it
  Chandra} imaging data in A2390, at roughly $70^{\prime\prime}$
distance from the cluster center. This brightness edge is marked by
the thick solid line in the upper-right panel of Fig. \ref{fig:xray},
clearly within the boundaries of our measured diffuse radio emission.

For the case of A2261, there is an extended structure to the
south-west, which appears to be uncorrelated to the region hosting the
radio halo emission. We therefore exclude this substructure from our
$\beta$-model fit. The remainder of the cluster emission is extremely
uniform within the central $\sim 1$ Mpc region.

Binning the XMM-Newton spectroscopic data using a Voroni tessellation
scheme indicates, apart from the prominent cool cores, a significant
azimuthal variation in the distribution of the hot gas, which can be
an indicator of past merger activity. Unfortunately, the uncertainties
in the tessellated temperatures in both data sets are large, and as
such the temperature images are not suitable for a more detailed
analysis of the cluster morphologies.

We conclude that while both fields show signs of residual structure
after removing a main component (as is expected at some level in any
hierarchical scenario), there are no typical signs of mergers in
either field. The strongest residual component in A2390 can correspond
to no more than a few per cent of the total gas mass in the system,
ruling out a major merger scenario, and the levels of substructure in
A2261 are lower still. We find no ripple-like features (which would
have indicated gas sloshing) from the available XMM-Newton data, but
we caution that the XMM resolution may be insufficient for detecting
sharp density features. A previously discovered brightness edge in
A2390 is located well inside the presently detected radio emission,
ruling it out as being directly related to the origin of the diffuse
emission.

\subsection{Cool cores in merging systems}

We find that the standard X-ray morphological indicators place these
two clusters apart from the parameter space quadrant of merging
systems, where most currently known radio halos reside. While the
X-ray brightness images indicate the presence of some substructure,
there are no clear signs of active merger activity. A possibility
would be that the cool cores have survived minor, non-disruptive
merger events, which nonetheless have created enough turbulent energy
to power radio halos.

The survival of cluster cool cores during mergers has been verified
through numerous simulations.  Early hydro-simulations of controlled
cluster mergers \citep{2006MNRAS.373..881P,2008MNRAS.391.1163P} found
that it is difficult to destroy cool-cores unless there is a head-on
collision involving a high mass ratio. More recent cosmological
hydro-simulations \citep{2015arXiv150904289H} came to similar
conclusions. 
Irrespective of the relative abundance of cool core/non-cool core
clusters, cool cores are destroyed only by low angular momentum major
mergers. Such a major merger scenario is not supported by the X-ray
analysis for either targets, but the occurrence of past minor mergers
cannot be ruled out. In the early simulations of
\cite{2008MNRAS.391.1163P} it was reported that off-axis mergers can
cause short-lived ``warm cores" with elevated entropy, which might be
the case for A2261. One main conclusion from these simulation results
is that the presence of cool-cores should not be taken as an indicator
of a relaxed, virialized systems. Observationally, prominent cool-core
clusters show a preference for a steady growth over $\sim 8$ Gyr
\citep{2013ApJ...774...23M}, during which time minor merger events
must have taken place. The two systems discussed in this Paper are
among the most massive galaxy clusters known ($M_{500} \gtrsim
10^{15}$ M$_{\odot}$), thus constituting fair targets for mergers over
their formation lifetime.

The question of whether a minor, off-axis merger can create enough
turbulent energy to cascade into particle acceleration is an open
issue. \cite{2009A&A...507..661B} pointed out that a decaying
turbulent kinetic energy following a merger strongly suppresses the
radio emission, which is a possible mechanism for creating a
bi-modality in the radio halo population. Several other factors,
including the magnetic field strength and a possible pre-existing
cosmic ray electron population, may play a role. If the two objects
discussed in this paper are considered as radio halos powered by
secondary (turbulent) CRe acceleration, they (together with the
similar object in CL1821$+$643) might indicate a new population of
radio sources found in massive cool-core clusters with decaying or
low-level turbulence, and may well represent an intermediate phase
between mini-halos and giant halos.

\subsection{A2390: An ultra-steep spectrum radio halo?}
\label{sec:a2390}

The diffuse radio source in A2390 presents an interesting case due to
its steep spectrum ($1.60 \pm 0.17$ at 20 cm). Such a steep spectrum
suggests a turbulent leptonic origin of its cosmic ray electrons, as
opposed to a hadronic origin that would require an unrealistically
large energy density for the relativistic protons
\citep{2008Natur.455..944B}. Indeed, the existence of {\it ultra-steep
  spectrum radio halos} is a distinguishing prediction of the
turbulent model, and only a handful of such steep-spectrum sources are
known (e.g. \citealt{2012A&ARv..20...54F} and references therein).
\cite{2014MNRAS.438..124Z} suggested a hybrid scenario in which giant
radio halos (GHRs) experience a transition from a hadronic emission
component in the center to a mainly leptonic component in the halo
outskirts. The relatively steep spectrum of A2390 suggests that this
may be a radio halo of leptonic origin in transition from a central
hadronic component.

Conversely, designating the A2390 radio emission as a mini-halo
(albeit with more than double the size of currently known mini-halos,
see \citealt{2014ApJ...781....9G}) with a turbulent origin of its
relativistic electrons would require Mpc-scale sloshing motion of its
gas core. As discussed in Section \ref{sec:xbright}, the X-ray
brightness of A2390 reveals an edge in the emission, thought to
originate from gas sloshing \citep{2015Ap&SS.359...61S}. However, the
presently measured extent of the radio emission extends well beyond
that brightness edge (Fig. \ref{fig:xray}, upper-right panel). The
simulations of \cite{2013ApJ...762...78Z} clearly demonstrate that
sloshing motion can generate turbulence only within the sloshing
fronts. Unless new sloshing-like features are found further outwards,
the radio emission in A2390 can be considered a giant radio halo
originating from merger-related turbulence.

\section{Summary and conclusions}
\label{sec:conclusions}

We have analyzed radio data for two clusters of galaxies hosting cool
cores, originating from a mass-selected sample containing some of the
most massive known clusters in the universe. Both systems have been
previously studied in the radio regime, however, we make the first
robust determination of their extended signals using radio images with
sufficient spatial dynamic range. In conjunction with the only
previously confirmed cool-core cluster hosting radio emission at this
scale \citep{2014MNRAS.444L..44B}, there are now three cases of giant
radio halos in cool-core cluster with regular X-ray morphologies.

The most important steps of our analysis, and the results obtained,
can be summarized as follows:

\begin{enumerate}
\item We calibrated JVLA (A2390), VLA (A2390, A2261) and GMRT (A2261)
  data to allow high dynamic range images of the targets at 1.4 GHz
  and, in the case of A2261, at 240 and 610 MHz.
\item We verified the recovery of extended emission and the robustness
  of the subtraction of compact components by means of simulations.
\item We determined the luminosities and spectral slopes of the
  extended signal components, applying a range of simulations to
  account for biases due to compact sources.
\item From the 1.4 GHz data, both systems were found to host radio
  emission on scales of about 1 Mpc. The low-frequency data is not
  deep enough for an accurate determination of the flux of the
  extended emission in A2261; however, the data are largely consistent
  with what would be expected from a radio halo of moderate spectral
  slope.
\item Scaling properties of the radio emissions with mass proxies were
  compared to known scaling relations. Given the substantial scatter
  in such relations, the current results are consistent with previous
  findings for radio halos.
\item We computed morphological measures quantifying the
  dynamical states of the clusters. In the space of power ratio versus
  concentration, the two targets lie in between typical radio halo
  clusters and typical ``radio quiet'' clusters. The centroid shift
  parameters indicate more relaxed morphologies than clusters
  typically associated with radio halos.
\item To test for density disturbances we fit and subtracted double
  $\beta$-models from the XMM-Newton X-ray photometric
  images. Although there are no obvious signs of ongoing mergers, we
  do find azimuthal variances in the gas distributions, in agreement
  with some previous studies.
\item The steep spectral slope of the A2390 radio emission is similar
  to some known ultra-steep spectrum radio halos, suggesting a
  possible leptonic origin in transition from a central hadronic
  component. The halo extends beyond a sloshing-like X-ray brightness
  discontinuity known from the {\it Chandra} data.

\end{enumerate}

Our main conclusion is that the two Mpc-scale diffuse radio emission
regions discussed in this work have properties similar to giant radio
halos, and should be considered as such. The fact that they are found
in two prominent cool-core clusters should not be surprising given
that these are extremely massive systems and thus fair targets for
minor, off-axis mergers during the formation lifetime of their cool
cores. 

We find some evidence for minor mergers based on X-ray brightness
images. Whether such non-disruptive mergers can generate enough
turbulent kinetic energy to eventually power particle acceleration,
and how long such elevated level of turbulence lasts, remain open
questions. Nevertheless, the current results indicate that the simple
picture of radio halos only occurring in actively merging clusters
with irregular morphologies is at best incomplete. Together with the
similar object in CL1821$+$643, the two objects discussed in this work
may be representative of an intermediate phase between mini-halos and
giant halos.

\section*{Acknowledgments}

MWS acknowledges partial support for this work from Transregio
Programme TR33 of the German Research Foundation (Deutsche
Forschungsgemeinschaft).
ABo acknowledges support by the research  group FOR 1254 funded by the
Deutsche Forschungsgemeinschaft:  ``Magnetisation of  interstellar and
intergalactic   media:   the    prospects   of   low-frequency   radio
observations".
ABa would like to acknowledge support from NSERC (Canada) through the
Discovery Grants program, as well as the Pauli Center for Theoretical
Studies UZH ETH.
The National Radio Astronomy Observatory is a facility of the National
Science Foundation operated under cooperative agreement by Associated
Universities, Inc. GMRT is run by the National Centre for Radio
Astrophysics of the Tata Institute of Fundamental Research.

\bibliographystyle{mn2e}   
\bibliography{cccp}{}    

\label{lastpage}

\end{document}